\documentclass[letter,12pt]{article}
\usepackage{jheppub} 
\usepackage{lineno}
\usepackage{amsmath,mathtools}
  \usepackage{graphicx}

\newcommand{\bea}{\begin{eqnarray}}
\newcommand{\eea}{\end{eqnarray}}

\newcommand{\CC}{\ensuremath{\mathcal C}}

\newcommand{\rmO}{\ensuremath{{\rm O}}}
\newcommand{\R}{\ensuremath{\mathbb R}}

\newcommand{\SL}{\ensuremath{{\rm SL}}}
\newcommand{\Sp}{\ensuremath{{\rm Sp}}}
\newcommand{\Sphere}{\mathbb{S}}
\newcommand{\SU}{\ensuremath{{\rm SU}}}
\newcommand{\U}{\ensuremath{{\mathrm U}}}
\newcommand{\Z}{\ensuremath{\mathbb Z}}

\usepackage{caption}
\usepackage{subfig}
\usepackage{graphicx}

\usepackage{ mathrsfs }
\usepackage{ upgreek }

\newcommand{\D}{\mathscr{D}}

\title{\boldmath TQFT gravity and ensemble holography}

\author{Anatoly Dymarsky}
\author{and Alfred Shapere}
\affiliation{Department of Physics and Astronomy, University of Kentucky,\\ 506 Library drive, Lexington, KY 40506}

\emailAdd{a.dymarsky@uky.edu}
\emailAdd{shapere@g.uky.edu}

\abstract{We outline a general derivation of  holographic duality between ``TQFT gravity'' -- the path integral of a 3d TQFT summed over different topologies -- and an ensemble of boundary 2d CFTs.  The key idea is to place the boundary ensemble on a Riemann surface of very high genus, where the duality trivializes.  The duality relation at finite genus is then obtained by genus reduction. 
Our derivation is generic and does not rely on an explicit form of the bulk or boundary  partition functions. It guarantees unitarity and    
suggests that the bulk sum should include all possible topologies. In the case of Abelian Chern-Simons theory with compact gauge group we argue that the weights of the boundary ensemble are equal, while the bulk sum reduces to a finite sum over equivalence classes of topologies,  represented by handlebodies with possible line defects.}

\begin{document}
\maketitle
\flushbottom

\section{Introduction}
\label{sec:intro}
The quest to understand quantum gravity, and the holographic correspondence when the bulk theory is dual to an ensemble of boundary QFTs,
has recently ignited interest in low-dimensional models of ``TQFT gravity.'' For an ensemble of two-dimensional QFTs, such a duality implies that the partition function averaged over the ensemble is equal to a sum of three-dimensional TQFT path integrals  evaluated on different topologies. 
Examples include  the 2d Ising model dual to the 3d Ising TQFT \cite{Castro:2011zq,Jian:2019ubz,Romaidis:2023zpx}, the ensemble of all Narain theories of central charge $c$ dual to ``$\U(1)$ gravity,'' i.e.~the perturbative part of  3d $U(1)^{c}\times U(1)^c$ Chern-Simons theory summed over handlebody topologies \cite{Afkhami-Jeddi:2020ezh,Maloney:2020nni,Datta:2021ftn}, and a handful of other examples involving ensembles of Narain theories \cite{cotler2020ads3,dymarsky2021comments,Benjamin:2021wzr,collier2022wormholes,Ashwinkumar:2021kav,Ashwinkumar:2023jtz,Ashwinkumar:2023ctt, Kames-King:2023fpa, Aharony:2023zit,  Forste:2024zjt} or more complicated RCFTs \cite{Castro:2011zq,Meruliya:2021utr}.  Studies of pure gravity in AdS${}_3$ in terms of Virasoro TQFT \cite{maloney2010quantum,giombi2008one,keller2015poincare,mikhaylov2018teichmuller,Chandra:2022bqq,Collier:2023cyw,collier2023solving,collier20243d,Chen:2024unp,Hung:2024gma,deBoer:2024kat} can also be included to this category. 

In the examples mentioned above, the duality is established by comparing explicit values of the bulk and boundary partition functions; however, a physical picture explaining such a correspondence is lacking. 
In certain cases, such as Virasoro TQFT \cite{collier20243d,collier2023solving},  the boundary ensemble is simply not well-defined. In other cases, e.g.~$\SU(2)_k \times {\SU(2)}_{-k}$ Chern-Simons  theory at certain levels $k$, the would-be boundary ensemble includes modular invariants with non-integral and/or negative weights and is therefore unphysical \cite{Meruliya:2021utr}.  These issues reflect the lack of a systematic understanding of the conditions under which an ensemble of CFTs admits a holographic interpretation. 

In this paper we  confirm and further develop the picture proposed in \cite{Barbar:2023ncl}, according to which the  boundary ensemble dual to a given TQFT includes all ``code CFTs'' corresponding to maximal gaugings of the bulk theory,\footnote{One can think of the ensemble of boundary theories as including all topological boundary conditions in the bulk, combining  TQFT blocks into  physically-consistent modular-invariant combinations.} while the bulk TQFT path integral is summed over various topologies
\bea
\sum_{{\rm CFTs}} \alpha_I\, Z_I =\sum_{\rm topologies}\  \beta_{\cal L}\, \Psi_{\cal L}\, . \label{duality}
\eea
We give a microscopic derivation of this duality relation %
for Abelian TQFTs and their corresponding ensembles, which from first principles fixes all weights $\alpha_I$ to be equal\footnote{Barring degeneracies, see the end of section \ref{sec:bulk}.} and the bulk sum  to include all  possible classes of 3d topologies, with possibly distinct weights $\beta_{\cal L}$. 
We then discuss generalizations of \eqref{duality}  to non-Abelian TQFTs and to higher dimensions.

We consider Abelian TQFTs in the next section.  After briefly reviewing Abelian 3d Chern-Simons theory, we illustrate the subtleties of determining both the ensemble weights $\alpha_I$ and the bulk topologies to sum over,  focusing on the example of $\Z_4$ gauge theory in section \ref{k=4}.  Then, in section \ref{sec:bulk}, we derive \eqref{duality} in the general case by using two representations of a projector on the space of all modular-invariant TQFT states.  Section \ref{sec:extensions} discusses various examples of Abelian Chern-Simons gravity, dual to ensembles of Narain theories, including the ``$U(1)$ gravity'' of \cite{Afkhami-Jeddi:2020ezh,Maloney:2020nni} dual to an average over the full Narain moduli space. We also discuss implications for the SymTFT picture, 
generalizations to non-Abelian theories, and theories in higher dimensions. The paper concludes with a discussion in section \ref{sec:discussion}.

\section{Abelian 3d TQFTs}

\subsection{Preliminaries: 3d Abelian Chern-Simons theory}
Our setup is the same as in  \cite{Barbar:2023ncl}: we consider 3d bosonic  anomaly-free $\U(1)^{2n}$ Chern-Simons (CS) theory 
\bea
\label{CS}
S={i\over 4\pi}\int K_{\alpha \beta}\, A^\alpha \wedge dA^\beta
\eea
with $K$ being the Gram matrix of an even lattice $\Lambda_0\subset \R^{n,n}$. The Wilson line operators $W_c(\upgamma)$ of this theory are labeled by elements of the discriminant group $c\in {\mathscr D}=\Lambda_0^*/\Lambda_0$ and 1-cycles $\upgamma$ \cite{Belov:2005ze}.

The Hilbert space ${\cal H}$ of this theory placed on a torus of genus 1 is of dimension ${\rm dim}\,{\cal H}=|{\D}|$. 
The  Hilbert space ${\cal H}^g$ over an arbitrary Riemann surface $\Sigma$ of genus $g$ is isomorphic to the tensor product
\bea
{\cal H}^g=\underbrace{{\cal H}\otimes \dots \otimes {\cal H}}_{g\, {\rm times}}.
\eea
This decomposition can be established explicitly using the orthonormal basis of wave-functions $|\Psi_{c_1\dots c_g}\rangle \in {\cal H}^g$ with $c_i \in \D$, defined as the path integral on the handlebody $M$ with $\partial M=\Sigma$,  with all $a$-cycles shrinkable inside $M$,
and with the Wilson line $W_{c_i}$ wrapping the  $i$th $b$-cycle. 
We provide explicit expressions in Appendix \ref{sec:CS}. 

The group $\D$ inherits a Lorentzian scalar product from $\Lambda_0^*$, which we denote by $\eta$. 
The maximal (Lagrangian) subgroups of the 1-form symmetry group, which 
correspond to the topological boundary conditions of this theory, are linear even self-dual codes \cite{Barbar:2023ncl} -- the subgroups $\CC \subset \D$ satisfying both evenness 
\bea
\forall\,  c\in \CC, \quad \eta(c,c)=0\, \,{\rm mod}\, \, 2,
\eea
(i.e., the anyons associated with $W_c$, $c\in \CC$  are bosons)
and self-duality, meaning that the only elements of $\D$ which have integral scalar product with all elements of $\CC$ are the elements of $\CC$ (i.e., the group of anyons with trivial mutual braiding is maximal). 

The topological boundary conditions on $\Sigma$ specified by $\CC$ define the wavefunction 
\bea
\label{codedecompose}
|\Psi_\CC\rangle^g=|\Psi_\CC\rangle \otimes \dots \otimes |\Psi_\CC\rangle=\sum_{c_i\in \CC} |\Psi_{c_1\dots c_g}\rangle,
\eea 
where $|\Psi_\CC\rangle$ is the wavefunction corresponding to $\CC$ when $\Sigma$ is a torus. 
We can think of $|\Psi_\CC\rangle^g$ as the wavefunction obtained by gauging a  maximal non-anomalous subgroup $\CC$ of the 1-form symmetry group.  In what follows we use the normalization $\langle \Psi_\CC|\Psi_\CC\rangle=|\CC|=|\D|^{1/2}$. For convenience we introduce a special notation for the vacuum state $|0\rangle^g\equiv |\Psi_{0\dots 0}\rangle$ produced by the path integral on a handlebody with shrinkable $a$-cycles and without any Wilson line insertions. 

Let us now consider the limit in which the genus-$g$ Riemann surface degenerates into a surface of lower genus $g-\tilde g$.  
This limit corresponds to taking the components of the imaginary part of the modular parameter $\Im(\Omega)$ associated with $\tilde g$ nonintersecting cycles to infinity, effectively reducing the genus from $g$ to $g-\tilde g$.  Then the only surviving wavefunctions  are $|\Psi_{c_1\dots c_g}\rangle$ with $c_i=0$, $g-\tilde{g} <i  \leq g$. 
Algebraically,  the effect of genus reduction, defined in terms of the Siegel $\Phi$ map \cite{RUNGE1996175}, can be written as 
\bea
\label{factorization}
\Phi:\,\, \, |\Psi\rangle^g \,\, \,   \rightarrow \,\, \,    {}^{\tilde g}\hspace{-1pt}\langle 0|\Psi\rangle^g=|\Psi\rangle^{g-\tilde g},
\eea
for any $|\Psi\rangle^g \in {\cal H}^g$. Since by linearity any code $\CC$ includes a unique zero element, we find that 
$\langle 0|\Psi_\CC\rangle=1$ and \eqref{codedecompose}  is compatible with $\Phi$.

\subsection{Holographic ensemble and sum over topologies}

The mapping class group $\mathscr{M}_g$ of a genus-$g$ Riemann surface $\Sigma$ 
 is the modular group 
 $\Sp(2g,\Z)$. For any $\gamma\in \Sp(2g,\Z)$, its action $U_\gamma: {\cal H}^g \rightarrow {\cal H}^g$  leaves all $|\Psi_{\CC}\rangle^g$ invariant. In fact, the set of $|\Psi_{\CC}\rangle^g$ for all even self-dual codes  (topological boundary conditions) $\CC$  spans the subspace of ${\cal H}^g$ invariant under modular transformations.

Thinking of $\Sigma$ as the boundary of a 3-geometry $M$, one can introduce ``coherent'' boundary conditions on $\Sigma$ by fixing certain holomorphic and antiholomorphic components of $A^i$ to be equal to $\xi^i,\bar \xi^i$ \cite{Aharony:2023zit}. These boundary conditions require introducing a complex structure on $\Sigma$ and explicitly depend on its modular parameter $\Omega$. 
The path integral  on $\Sigma \times \R^+$ with these boundary conditions defines  
a basis state $|\xi, \bar \xi\rangle$, and the wavefunction of $|\Psi\rangle^g \in {\cal H}^g$ 
written in this basis is
\bea
\Psi(\Omega, \xi, \bar \xi):=\langle \xi, \bar \xi|\Psi\rangle^g,\qquad \Psi(\gamma\, \Omega, \gamma\, \xi, \gamma\, \bar \xi)=\langle \xi, \bar \xi| U^\dagger_\gamma|\Psi\rangle^g
\eea
where $\gamma\, \Omega$ denotes the action of $\gamma$ on $\Omega$.  
The wavefunction of $|\Psi_{\CC}\rangle^g$ written in this basis 
is  the modular-invariant partition function (more accurately, path integral \cite{Kraus:2006nb,Aharony:2023zit}) of the Narain code CFT associated with $\CC$ \eqref{codedecompose}
\bea
\label{basis}
\langle \xi, \bar \xi|\Psi_\CC\rangle^g=\Psi_\CC(\Omega, \xi,\bar \xi)=Z_\CC(\Omega,\xi,\bar \xi).
\eea 
The path integral $Z_\CC(\Omega,\xi,\bar \xi)$ is defined in terms of the even self-dual Lorentzian lattice 
\bea
\label{lambdac}
\Lambda_\CC =\sum_{c\in \CC} \Lambda_c,\quad \Lambda_c\equiv\{c+u\, |\, u\in \Lambda_0\},
\eea
which is the Construction A lattice associated with the code $\CC$. 
The possible choices of coherent boundary conditions are parameterized by the Narain moduli space 
${\cal M}_n={\rmO}(n,n;\Z)\backslash {\rmO}(n,n)/{\rmO(n)}\times \rmO(n)$. Different  choices correspond to inequivalent Euclidean scalar products on $\Lambda_0^*$.  We call a particular choice of boundary conditions an ``embedding''  \cite{angelinos2022optimal,Aharony:2023zit}.

Following \cite{Barbar:2023ncl} and based on the ideas and results of \cite{Dymarsky:2020qom,Benini:2022hzx,Henriksson:2022dml,Aharony:2023zit}, we consider averaging over the  ensemble of all maximal gaugings of 1-form symmetries (topological boundary conditions) in the TQFT \eqref{CS}, which gives rise to an ensemble of code CFTs.  The partition function averaged over this ensemble  with weights $\alpha_I$ is then
\bea
\label{ensemble}
\langle Z\rangle=\sum_I \alpha_I\, Z_{\CC_I}.
\eea
This expression is manifestly modular invariant and can therefore be expressed as a Poincar\'e series
\bea
\label{PS}
\langle Z\rangle\propto  \sum_{\gamma \in \Gamma_s\backslash \Sp(2g,\Z)} \Psi_{\rm seed}(\gamma\, \Omega, \gamma\, \xi, \gamma\, \bar \xi)
\eea
with some appropriate seed wavefunction (here $\Gamma_s\subset \Sp(2g,\Z)$ is the stabilizer subgroup of $ \Psi_{\rm seed}$). When the seed is particularly simple, say, the vacuum character -- i.e.~the wavefunction of the TQFT on the handlebody geometry $\Psi_{\rm seed}=\langle \xi,\bar \xi|\Psi_{0\dots 0}\rangle^g$ -- the Poincar\'e series can be interpreted as the TQFT path integral summed over a class of 3d topologies, pointing towards a holographic interpretation of \eqref{PS} in terms of ``TQFT gravity.'' 

This is the case, for example, for
$AB$ Chern-Simons theory, 
\bea
\label{AB}
S={i k\over 2\pi}\sum_{i=1}^n \int A_i\wedge dB_i,
\eea
with prime or square-free level $k$, as is verified for arbitrary genus  in \cite{DHP}. In this case all Narain theories in the ensemble have equal weights $\alpha_I$, as follows from the transitive action on the ensemble of the group $\rmO(n,n,\Z_k)$  preserving the scalar product $\eta$  \cite{Aharony:2023zit}. 
Equivalently,  the ensemble of Narain code CFTs in this case forms an orbifold groupoid with global $(\Z_k)^n$ symmetry, admitting a transitive action of  $\rmO(n,n,\Z_k)$  \cite{Gaiotto:2020iye}.

To illustrate the above ideas we briefly discuss the case of level $k=2$, for which the TQFT \eqref{AB} corresponds to $n$ copies of the toric code \cite{Shao:2023gho}. The topological boundary conditions in this case are described by binary even self-dual codes of length $2n$ with scalar product 
\bea
\eta(c,c')=(\vec{a}\cdot  \vec{b}'+ \vec{a}'\cdot  \vec{b})/k,\qquad c=(\vec{a},\vec{b}),\ c'=(\vec{a}',\vec{b'})\in \D= \Z_k^{2n}.
\eea
These codes were discussed in detail in \cite{Dymarsky:2020qom}. For $n=1$ there are only two even self-dual codes, $\CC_1=\{(a,0)\}$ and  $\CC_2=\{(0,b)\}$, with $a,b\in \Z_2$, which correspond to Dirichlet and Neumann topological boundary conditions. The group $\rmO(1,1,\Z_2)=\Z_2$ (electro-magnetic duality of $\Z_2$ gauge theory) acts on the set of codes transitively, by exchanging $c=(a,b)\leftrightarrow c'=(b,a)$. The Narain theories corresponding to $\CC_1$ and $\CC_2$ are compact scalars of radii $R_1=\sqrt{2}r$ and  $R_2=r/\sqrt{2}$,
\bea
\label{Z=Psi}
Z_{R_I}(\Omega,\xi, \bar \xi)=\langle \xi,\bar \xi|\Psi_{\CC_I}\rangle^g,
\eea
where $r$ is an arbitrary embedding parameter. The Poincar\'e series in this case is also simple. At genus $g=1$ it reduces to three terms, generated by representatives $1,S,TS$ of the coset $\Gamma_0(2)\backslash \SL_2(\Z)$  (we have suppressed $\xi,\bar\xi$ for simplicity).  
Thus \eqref{PS} takes the form
\bea
\label{TCP}
Z_{R_1}+Z_{R_2}= \Psi_0(\tau)+\Psi_0(-1/\tau)+\Psi_0(-1/(\tau+1)).
\eea
Here $\Psi_0(\tau,\xi, \bar \xi)=\langle \xi,\bar \xi|\Psi_0\rangle$ is equal to the theta-series of $\Lambda_0$ (mapped into $\R^{2}$ with embedding parameter $r$) divided by the absolute value squared of the  Dedekind  eta function \cite{Aharony:2023zit}. 
The generalization of this expression to higher genus is straightforward, because the transitive action of $\rmO(n,n,\Z_2)$ on codes fixes all the $\alpha_I$ to be equal and implies that the  vacuum character can be chosen as the seed $\Psi_{\rm seed}=\Psi_{0\dots 0}$ \cite{Aharony:2023zit,DHP}. 

To close this section, we comment on the large-$p$ limit. In this case, when $n>2g$ the boundary ensemble (apparently) becomes the ensemble of all Narain theories weighted with the Haar measure, while the Poincar\'e series in the bulk  becomes a real Eisenstein series  \cite{Aharony:2023zit}, evaluating the partition function of \U(1) gravity \cite{Afkhami-Jeddi:2020ezh,Maloney:2020nni}.  

\subsection{Non-decoupling wormholes and singular topologies}
\label{k=4}
The simplest scenario,  in which the Poincar\'e series of the vacuum character $\Psi_{0\dots 0}$ -- the path integral of the TQFT summed over handlebody topologies -- can be interpreted as a boundary CFT ensemble, is not universal. As an example let us consider $\Z_4$ gauge theory, i.e. the $k=4$ $AB$ theory \eqref{AB} with $n=1$.  
There are three possible topological boundary conditions for this theory, corresponding to the even self-dual codes $\CC_1=\{(a,0)\}, \quad \CC_2=\{(0,b)\},\quad \CC_3=\{(2a,2b)\}$. The first two are exchanged by $\rmO(1,1,\Z_4)=\Z_2$ ``electro-magnetic'' duality \cite{Gaiotto:2020iye} and the third code is invariant. The corresponding  Narain CFTs are compact scalars of radii  $R_1=2r, R_2=r/2$, and $R_3=r$, where $r$ is an arbitrary embedding parameter.  As in \eqref{Z=Psi},  $Z_{R_I}=\Psi_{\CC_I}$. 

At genus $g=1$, the modular group acting on $\cal H$ reduces to $\Gamma(4)\backslash \SL_2(\Z)=\SL_2(\Z_4)$,  while the ``vacuum character'' $\Psi_{0}(\tau,\xi,\bar \xi)$ is stabilized by $\Gamma_0(4)$. The quotient $\Gamma_0(4) \backslash \SL_2(\Z)$ consists of $6$ equivalence classes, with representatives $1$, $T^\ell S$  for $0\leq \ell \leq 3$, and $S T^2 S$.  The Poincar\'e series can be decomposed into a sum of code CFTs as  follows
\bea
\label{g=1}
Z^{g=1}_{2r}+Z^{g=1}_{r/2}+{1\over 2}Z^{g=1}_r= \sum_{\gamma \in \Gamma_0(4) \backslash \SL_2(\Z)}  U_\gamma\,  \Psi_{0},
\eea
where with a small abuse of notation we use $U_\gamma$ to denote the representation of the modular group acting on $\Psi_c(\tau,\xi,\bar \xi)$.
At genus $g=2$ the ``vacuum character'' $\Psi_{00}$ is stabilized by the congruence subgroup
$$
\Gamma_0^{(2)}(4)=\left\{
\begin{pmatrix}
A& B\\ C& D
\end{pmatrix}
: C\equiv 0\ {\rm (mod~4)}
\right\}
\subset \Sp(4,\Z).
$$
The genus-2 version of \eqref{g=1} is
\bea
\label{g=2}
Z^{g=2}_{2r}+Z^{g=2}_{r/2}+{3\over 4}Z^{g=2}_r= {1\over 6} \sum_{\gamma \in \Gamma_0^{(2)}(4) \backslash \Sp(4,\Z)} U_\gamma\, \Psi_{00}.
\eea
where the Poincar\'e series on the right side has 120 terms.  
Comparing \eqref{g=1} and \eqref{g=2}, we find  that the coefficients $\alpha_I$ are $g$-dependent, which  conflicts with the ensemble average interpretation. The latter requires that once the weights $\alpha_I$ are specified, the ensemble of CFTs can be placed on any 2d geometry. 

To trace the origin of the genus dependence of the $\alpha_I$, we start with \eqref{g=2} and apply
the genus-reducing map of 
\eqref{factorization}, first factorizing the genus-2 Riemann surface into two tori $T$ and $T'$ and then taking $\tau'\rightarrow i\infty$.
The resulting genus-1 equality is
\bea
\label{g1}
Z^{g=1}_{2r}+Z^{g=1}_{r/2}+{3\over 4}Z^{g=1}_r= \sum_{\gamma \in \Gamma_0(4) \backslash \SL_2(\Z)} U_\gamma\, \left[ \Psi_{0}+ {1\over 24}  \sum_{a,b=0,1} \Psi_{(2a,2b)}\right].
\eea
This is consistent with \eqref{g=1} because $Z^{g=1}_{r}=  \sum_{a,b=0,1} \Psi_{(2a,2b)}$.

\begin{figure}
\subfloat[Wormhole with a surface defect]{\includegraphics[width=.48\textwidth]{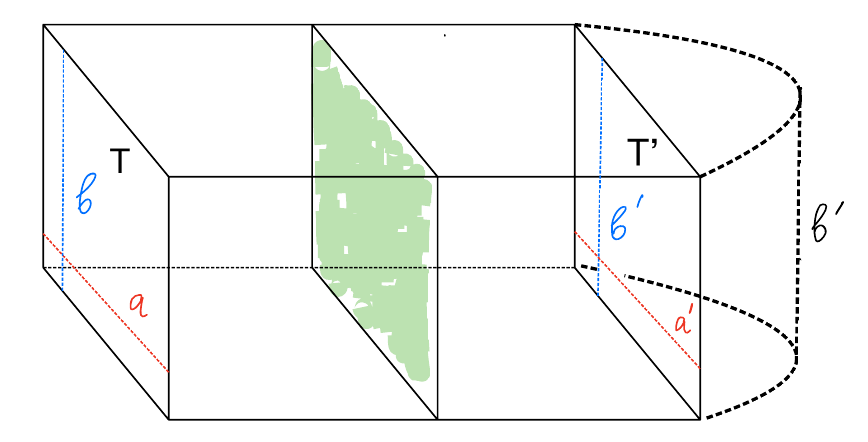}}
\subfloat[Handlebody  $M$ with a line defect]{\includegraphics[width=0.43\textwidth]{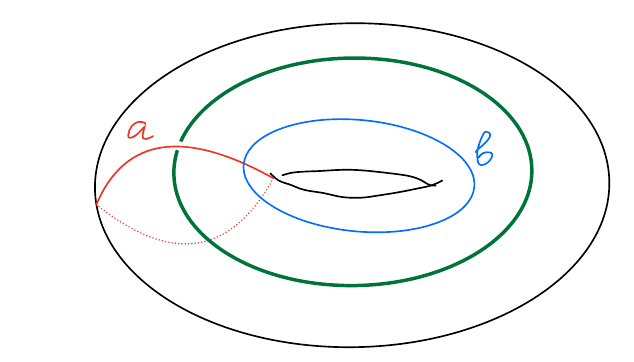}}
\caption{a). Wormhole geometry with a surface defect implementing $a=2a'$, $2b=b'$ gluing of $T$ and $T'$ tori.   b). Resulting handlebody geometry with a line defect after $a'$ is shrunk inside $T'$.  Now both $a,b$ are non-shrinkable.}
\label{fig:wormhole}
\end{figure}

The appearance of an extra term in the right-hand side of  \eqref{g1}, in addition to the  Poincar\'e series of the vacuum character, is due to  a ``non-decoupling wormhole.'' These are singular wormhole geometries connecting the two tori into which the original genus-2 Riemann surface has factorized, with one of the tori subsequently degenerating into a cylinder. Since the theory in the bulk is topological, contributions of these wormholes are not suppressed in comparison with the factorized geometries. 

One specific topology (among many) giving rise to the last term \eqref{g1} can be constructed as follows. Consider a 3d cylinder composed of the torus  $T$ with cycles $a,b$ multiplied by an interval $I$, along with a cylinder $T'\times I$ whose cross-section is the torus $T'$ with cycles $a',b'$. Then glue together one of the boundaries of each cylinder, such that cycle $a'$ covers $a$ twice, and similarly such that $b$ covers $b'$ twice. This is shown in panel (a) of Fig.~\ref{fig:wormhole}. 
In terms of the sum on the RHS of \eqref{g=2}, it emerges from a smooth handlebody $M$ ending on a genus $g=2$ Riemann surface $\Sigma$, after an appropriate mapping class group transformation $U_\gamma$ (still a smooth geometry), where e.g.
\begin{eqnarray}
\gamma=\left(
\begin{array}{cccc}
 2 & 0 & 0 & -1 \\
 1 & 0 & 0 & 0 \\
 0 & 1 & 0 & 0 \\
 0 & -2 & 1 & 0 \\
\end{array}
\right),
\end{eqnarray}
 which is subsequently factorized such that $\Sigma\rightarrow T \times T'$.  
The resulting singular geometry may be viewed as a 3d cylinder with an inserted  surface defect. In $k=4$ $AB$ theory this defect becomes a Kapustin-Saulina non-invertible surface operator  \cite{Kapustin:2010hk,Kapustin:2010if} (condensation defect of \cite{Roumpedakis:2022aik}), imposing topological boundary conditions defined by $\CC_3$, i.e.~the surface operator is a projector onto $|\Psi_{\CC_3}\rangle$.\footnote{A wormhole with a  Kapustin-Saulina surface operator  was recently studied in \cite{Raeymaekers:2023ras} in the context of $\U(1)_4 \times \U(1)_{-4}$ Chern-Simons theory. The logic there is complementary to ours: such wormholes were required to match a particular boundary ensemble. Here we emphasize that such geometries  inevitably emerge from smooth geometries via genus reduction.} Finally, the other boundary of the $T'$ cylinder is glued to a handlebody with a shrinkable $a'$ cycle.  This implements ${}^1 \langle 0|U_\gamma|0\rangle^2$.
Alternatively,   instead of gluing a handlebody, one can send $\tau'\to i\infty$, with $T'$ degenerating into a circle.  The resulting geometry is a handlebody ending on $T$, with a line defect along the $b$ cycle, shown in green in panel (b) of Fig.~\ref{fig:wormhole}.  The cycle $a$ is not shrinkable in this geometry, but any double cover, i.e.~any Wilson line around $a$  with a charge $c\in \D$ divisible by $2$, is trivial as it can be extended into $T'$ where it is  shrinkable. Similarly, any Wilson  line of charge $c$ around $b$ becomes a Wilson line of charge $2c$ around $b'$.  In the case of $k=4$ $AB$ theory, the above construction defines a  geometry $M$ with homology  group $H_1({M},\Z_4)=\Z_2\times \Z_2$. In the end, the line defect is equivalent to a sum of Wilson line operators with charges $c\in \D$ divisible by $2$. 
In other theories the fate of surface defect could be different. 
Say, for odd primes $k=p$, all elements of $\D$ are divisible by $2$ and therefore the surface defect dissolves, i.e.~it becomes an identity operator, effectively resulting in the original handlebody topology. The case of $k=2$ is somewhat different. In this case all Wilson lines wrapping the $a$ cycle are non-trivial, but all lines wrapping the $b$ cycle automatically have even charge and therefore trivial. This means the $b$ cycle is  effectively  shrinkable,  i.e.~the surface defect, which in this case is not  a condensation defect of \cite{Roumpedakis:2022aik},  implements the $S$ modular transformation, mapping $a$ to $-b$. The resulting topology is that of  a handlebody, which is consistent with the RHS of \eqref{TCP} being a sum over equivalence classes of handlebody topologies. 

The genus reduction procedure also creates many other smooth non-handlebody geometries $M$, with one cycle of its boundary ${\partial M}=T$ shrinkable in $M$. The Abelian TQFT path integral on such a geometry, up to an overall  factor, is the same as on a handlebody geometry with the same shrinkable cycle $\upgamma$ in $T$. The contribution of these non-handlebody geometries effectively renormalizes the overall coefficient in front of the Poincar\'e series. 

The presence of non-decoupling wormholes  suggests that the ``sum over topologies'' in the bulk, in addition to the Poincar\'e series of the vacuum character, should in general  include other states.  This raises the questions of which singular geometries and corresponding states to include, whether there is a physically-preferable way to choose the weights $\alpha_I$, and whether the  resulting bulk sum admits a ``gravitational'' interpretation. In the remainder of this section we discuss to what extent the first two questions can be answered in the $k=4$ AB theory, based on the intrinsic mathematical structure of the problem. We will skip most of the technical details, although some can be found in Appendix \ref{genusreduction}. 

Two of the key components of this mathematical structure are the group of  modular transformations $\Sp(2g,\Z_k)$  and the group $\rmO(n,n,\Z_k)$  that reshuffles the set of codes. They form a reductive dual pair within the bigger group $\Sp(4gn,\Z_k)$,  meaning that $\Sp(2g,\Z_k)$ is the centralizer of $\rmO(n,n,\Z_k)$ in $\Sp(4gn,\Z_k)$ and vice versa; $\Sp(4gn,\Z_k)$ is the automorphism group of ${\cal W}$, the group of Wilson line operators  \cite{DHP}. 
The code-based states $|\Psi_{\CC_I}\rangle$ form a  basis for the  modular invariant subspace of $\cal H$.   Focusing on $g=1$, the ``topologies'' are defined algebraically by Lagrangian subgroups ${\cal L}\subset H_1(T,\Z_k)=\Z_k \oplus \Z_k$,  which are
self-dual with respect to the symplectic product   \cite{Barbar:2023ncl}
\bea
\omega=\left(
\begin{array}{cc}
0 & -1\\
1 & 0
\end{array}
\right).
\eea
The corresponding states $|\Psi_{\cal L}\rangle$, which can be understood as stabilizer states of  quantum codes specified by  symplectic codes $\cal L$ \cite{Barbar:2023ncl},\footnote{For example, the ``vacuum'' state 
$|0\rangle=|\Psi_{{\cal L}_0}\rangle$ is a stabilizer state defined by 
 ${\cal L}_0=(a,0)\in \Z_k \oplus \Z_k$; see also \cite{Salton:2016qpp}.} 
form a basis for the subspace of $\cal H$ invariant under  $\rmO(n,n,\Z_k)$. If $k$ is prime or square-free,   $\rmO(n,n,\Z_k)$ acts transitively on codes and the modular group acts transitively on ``topologies'' $\cal L$, fixing all weights $\alpha_I$ to be the same and all handlebody topologies (there are no singular topologies in this case) to enter the bulk sum  with equal coefficients.  There is a unique state in $\cal H$ invariant under both groups, which ensures that averaging $|\Psi_\CC\rangle$ over all codes is proportional to the  Poincar\'e series of $|0\rangle$ \cite{DHP} as, for example, in \eqref{TCP}. 

When $k=4$, there are $7$  ``topologies'' $\cal L$, with the corresponding states
\bea
\label{topologies}
|0\rangle, \quad T^\ell S|0\rangle\ (0\leq \ell\leq 3),\quad ST^2 S|0\rangle, \quad {1\over 2^n}\sum_{a,b=\Z_2^n} |\Psi_{(2a,2b)}\rangle
\eea
which are exactly the states that can emerge in the degeneration limit from the TQFT path integral over a  handlebody geometry bounding a  higher genus  Riemann surface.

The first six states in \eqref{topologies} correspond to nonsingular handlebodies,; they form an orbit under the modular group. The last state corresponds to a handlebody with specific Wilson loop insertions; it is  modular-invariant by itself.  This means that there are two states in $\cal H$ invariant under both $\rmO(n,n,\Z_4)$ and $\SL_2(\Z_4)$. Thus, even if we   postulate invariance under $\rmO(n,n,\Z_4)$  (we provide physics arguments to  justify doing that in next section), there is no unique mathematically preferable choice of  weights $\alpha_I$. Rather, for  any $n$ there is a one-parameter family of choices, which for $n=1$ can be  written explicitly  as 
\bea
\label{id4}
Z^{g=1}_{2r}+Z^{g=1}_{r/2}+\alpha\,  Z^{g=1}_r= \sum_{\gamma \in \Gamma_0(4) \backslash \SL_2(\Z)} U_\gamma\, \Psi_{0}+ (\alpha-\frac12\,) \sum_{a,b=0,1} \Psi_{(2a,2b)},
\eea
See Appendix \ref{genusreduction} for additional details. 

For general values of $n$ and $g$, 
there will be a total of $(n+1)$ distinct orbits of $\rmO(n,n,\Z_4)$ and $(g+1)$ orbits of $\Sp(2g,\Z)$.   These orbits consist of even self-dual codes $\CC$ or  self-dual symplectic codes $\cal L$,  each of which are  Lagrangian subgroups of $\Z_4^{2J}$ with $J=n$ or $J=g$ respectively,  and which are isomorphic to
\bea
\Z_4^{j}\times (\Z_2\times \Z_2)^{J-j},\qquad 0\le j \le J.
\eea
The $J+1$ distinct orbits are labeled by $j$, which runs from $0$ to $J$. 
The number of independent parameters in the generalization of \eqref{id4} is therefore $\min (n,g)$.\footnote{For general $k$  the number of orbits invariant under both groups increases 
polynomially with $\min (n,g)$.  In particular, for $k=p^l$ and $n=1$, the number of orbits of $\rmO(1,1,\Z_{p^l})$ is equal to $P(l)$, the number of partitions of $l$. For $n>1$, the orbits of $\rmO(n,n,\Z_{p^l})$ correspond to bounded partitions of $nl$ into positive integers $\le l$. The total number of orbits for general $k$ is determined multiplicatively; it grows polynomially with $n$ and subpolynomially with $k$.}


We conclude this section by reiterating that there is apparently no mathematically  unique preferred choice of $\alpha$ in \eqref{id4},  
which is more fundamental than other choices. 
As we will see below, this is because different choices of ensemble weights $\alpha_I$ arise in different physical contexts.



\subsection{Bulk derivation}
\label{sec:bulk}
We now aim to provide a derivation of \eqref{duality},  universal for any anomaly-free Abelian Chern-Simons theory, which will fix from first principles the weights $\alpha_I$ and the classes of topologies appearing in the bulk sum. We first note that the equality  \eqref{duality} can be understood as a purely bulk statement, i.e., as a relation between different states of the 3d TQFT.  Specific examples involving Narain theories, e.g.~\eqref{TCP}, can be obtained using the coherent state basis \eqref{basis} for the Hilbert space. 

The key step is to consider a genus-$g$ density matrix 
\bea
\label{rho}
\rho_g=\sum_I \alpha_I |\psi_I \rangle \hspace{-1pt}\langle \psi_I|, 
\eea
where $|\psi_I\rangle=|\Psi_{\CC_I}\rangle^g /|\D|^{g/2}$ is a normalized genus-$g$ state created by the topological boundary conditions specified by the code $\CC_I$.    From the point of view of the TQFT, the projector $|\psi_I \rangle \langle \psi_I|$ is  proportional to a Kapustin-Saulina surface operator \cite{Kapustin:2010hk,Kapustin:2010if}. It can also be interpreted in quantum information theoretic terms as a projector onto a stabilizer state associated with a CSS quantum stabilizer code, defined in terms of the classical code $\CC_I$. We develop this picture in \cite{BDS}.  

The density matrix \eqref{rho} can be understood as a linear combination of surface operators. 
This definition is consistent with genus reduction
\bea
\label{trace}
\rho_{g-g'}={\rm Tr}_{g'} \rho_g,
\eea
where the partial trace is taken over a genus-$g'$ subspace of the full Hilbert space. (We note that genus reduction by partial trace is {\it a priori} different from the factorization map we discussed previously.)  For any finite $g$, the states $|\psi_I\rangle$ are not orthogonal; their scalar product is proportional to the number of common $g$-tuples of codewords in the corresponding codes,
\bea
\label{metric}
g_{IJ}=\langle \psi_I |\psi_J\rangle={|\langle \Psi_{\CC_I}|\Psi_{\CC_J}\rangle|^g\over |\D|^g}={|\CC_I \cap \CC_J|^g\over|\D|^g}.
\eea
Given that for distinct codes $|C_I \cap C_J|$ is a proper divisor of $|\D|$, 
in the limit $g\rightarrow \infty$ all vectors become mutually orthogonal and form an orthonormal basis for the subspace of ${\cal H}^g$ invariant under the modular group $\Sp(2g,\Z)$.  With the choice $\alpha_I=1$  for all $\CC_I$,  $\rho$ becomes a projector on the modular-invariant subspace, 
\bea
\label{bulk}
\rho_g=\sum_I |\psi_I \rangle \langle \psi_I|={1\over |\mathscr{M}_g|}\sum_{\gamma\in \mathscr{M}_g} U_\gamma,\qquad g\rightarrow \infty.
\eea
This is the main identity, valid only in the large-$g$ limit, from which \eqref{duality} follows by genus reduction.\footnote{One can start with a finite $g$ version of \eqref{bulk}, with the LHS becoming  $\sum_{IJ} |\psi_I\rangle  g^{IJ} \langle \psi_J|$ and the RHS remaining the same. Existence of a smooth $g\rightarrow \infty$ limit follows from the form of $g_{IJ}=\delta_{IJ}+O(2^{-g})$ as follows from \eqref{metric}.}   

There are two distinct ways to perform a reduction from genus $g+\tilde g$ to $g$ -- by taking a partial trace as in \eqref{trace} or by applying the Siegel map \eqref{factorization} -- which become equal in the large-$\tilde{g}$ limit,
\bea
\nonumber
\rho_g=\sum_I |\psi_I \rangle \langle \psi_I|= \lim_{\tilde g \rightarrow \infty}  {1\over |\mathscr{M}_{g+\tilde g}|}\sum_{\gamma\in \mathscr{M}_{g+\tilde g}} {\rm Tr}_{\tilde g}\,  U_\gamma = \lim_{\tilde g \rightarrow \infty}  {|\D|^{\tilde g}\over |\mathscr{M}_{g+\tilde g}|}\sum_{\gamma\in \mathscr{M}_{g+\tilde g}}{}^{\tilde{g}}{\langle} 0| U_\gamma |0\rangle^{\tilde g}. \\
\label{bulkg}
\eea 
The holographic identity \eqref{duality}, with all $\alpha_I=1$, then takes the form
\bea
\label{PsiBulk}
\sum_I |\Psi_{\CC_I}\rangle&=&\Psi_{\rm bulk},
\eea
where the wavefunction of TQFT quantum gravity is obtained by acting with $\rho_g$ on $|\D|^g |0\rangle^g$,
\bea
\label{finitegbulk}
\Psi_{\rm bulk}&=&  \lim_{\tilde g \rightarrow \infty}  {1\over |\mathscr{M}_{g+\tilde g}|}\sum_{\gamma\in M_{g+\tilde g}}\left( {\rm Tr}_{\tilde g}\,  U_\gamma\right)|0\rangle^g= \lim_{\tilde g \rightarrow \infty}  {|\D|^{\tilde g}\over |\mathscr{M}_{g+\tilde g}|}\sum_{\gamma\in \mathscr{M}_{g+\tilde g}} {}^{\tilde g} \langle 0|U_\gamma|0\rangle^{g+\tilde{g}}.\quad 
\eea

We would like to emphasize two features implicit in our proposal. First,  $\Psi_{\rm bulk}$, which is the 
analog of the Hartle-Hawking wavefunction \cite{hartle1976path}, is automatically invariant under all symmetries in the bulk -- specifically, under  the action of the invertible surface operators of the original TQFT, which act on $\Psi_{\rm bulk}$ trivially.  This means that all global symmetries in the bulk are gauged, 
consistent with the expectation that quantum gravity  admits no global symmetries \cite{Banks:2010zn,Harlow:2018tng}. 
Technically,  invariance under invertible surface operators follows from the fact that the weights of the states $\Psi_{\CC_I}$ mapped into each other by such operators are all the same.\footnote{The same technical point was recently given an opposite interpretation in  \cite{Ashwinkumar:2023jtz}, as a mechanism for a theory of ``quantum gravity'' to admit  global symmetry acting on the ensemble of boundary CFTs. } As we pointed out  in the case of $k=4$ AB theory above, this invariance by itself is not always sufficient to fix the weights $\alpha_I$ or the form of $\Psi_{\rm bulk}$.

\begin{figure}
\subfloat[Modular transformation]{\includegraphics[width=.48\textwidth]{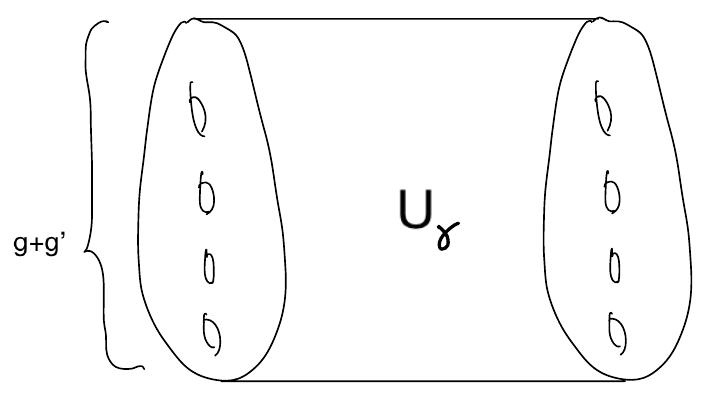}}
\subfloat[Wormhole geometry from genus reduction]{\includegraphics[width=0.53\textwidth]{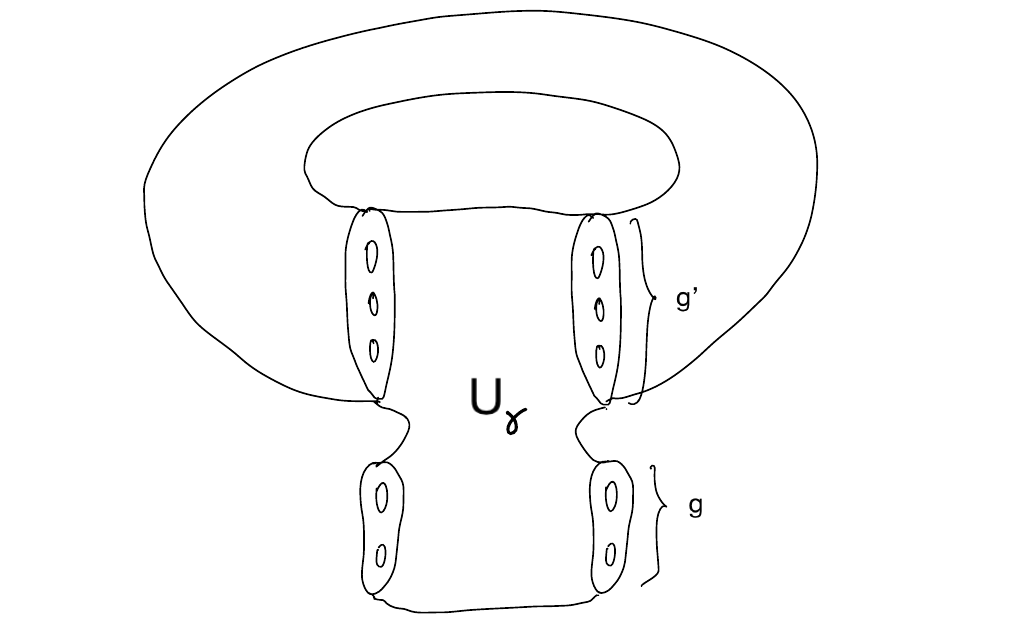}}
\caption{(a) 3d geometry implementing the modular transformation $U_\gamma$ on a Riemann surface of genus $g+g'$.  (b) The resulting geometry after factorization and partial trace, implementing ${\rm Tr}_{g'} U_\gamma$.}
\label{fig:trace}
\end{figure}

The second point concerns the classes of topologies contributing to $\Psi_{\rm bulk}$.  Starting from a high genus $g+\tilde{g}$, the operator $U_\gamma$ acting on ${\cal H}^{g+\tilde{g}}$ can be visualized as a 3-manifold with two boundaries, each being a Riemann surface of genus $g+\tilde{g}$.  The partial trace of $U_\gamma$ in \eqref{PsiBulk} creates wormhole topologies, as visualized in Fig.~\ref{fig:trace}. This means that the ``sum over topologies'' contributing to the genus-$g$ wavefunction $\Psi_{\rm bulk}$ is not limited to handlebody topologies, but includes wormhole geometries of arbitrarily complex topology. As we have seen above, these geometries may become singular after genus reduction.  In the limit $\tilde{g}\rightarrow \infty$ we expect that all possible orientable 3d topologies\footnote{It is easy to see that ${}^g\langle 0|U_\gamma|0\rangle^g$ creates all possible closed 3d topologies in the limit $g\to \infty$.  Indeed, the corresponding topologies are precisely the set of all Heegaard splitings of genus $g$ \cite{heegaard}.} can be constructed this way and will, at least in principle, contribute to the wavefunction of quantum gravity. 

To summarize, we have outlined a general derivation of holographic duality \eqref{duality}  for any Abelian anomaly-free Chern-Simons theory. The duality is understood as a purely bulk statement, an equality between different superpositions of states in the TQFT Hilbert space ${\cal H}^g$.  The main idea  is to start at infinite genus $g$, when the equality trivializes to \eqref{bulk}, and to  subsequently obtain the finite-$g$ equality via genus reduction. In principle all possible topologies, including singular ones obtained via genus reduction,  contribute at any finite $g$. 

In the Abelian case  the sum over topologies necessarily simplifies. Compact Abelian Chern-Simons theory 
at finite level can distinguish only a finite number of equivalence  classes of different topologies,  with the representatives being either pure handlebody topologies or handlebody topologies with particular line insertions. We saw an example of the latter in the previous subsection; more details can be found in Appendix \ref{genusreduction}.

For any given  $g+\tilde{g}$, the representation of the mapping class group  $\mathscr{M}_{g+\tilde{g}}$ on equivalence classes of topologies is finite in the  Abelian case. Therefore, the reduction \eqref{finitegbulk} to a given genus $g$   produces a $\tilde{g}$-independent number of terms, with $\tilde{g}$-dependent coefficients and with a well-defined $\tilde{g}\rightarrow \infty$ limit. Thus for $n$ copies of $k=4$ AB theory the reduction of the Poincare series of $\Psi_0^{\tilde g}=\Psi_{0\dots 0}$ from genus $1+\tilde{g}$ to $g=1$ would yield 
\bea
{|\D|^{\tilde{g}}\over |\mathscr{M}_{(1+\tilde{g})}|}  \sum_{\gamma \in \Gamma^{\tilde{g}+1}_0(4) \backslash Sp(2\tilde{g}+2,\Z)} 
{}^{\tilde{g}}\langle 0|U_\gamma |0\rangle^{\tilde{g}+1}=
a_n \sum_{\gamma \in \Gamma_0(4) \backslash \SL_2(\Z)} U_\gamma\, |0\rangle+ {b_n\over 2^n}  \sum_{a,b=0,1} |\Psi_{(2a,2b)}\rangle, \nonumber
\eea
with $\tilde{g}$-dependent $a_n,b_n$. Taking the $\tilde{g}\rightarrow \infty$ limit  gives 
\bea
\label{k4P}
\sum_{I=1}^{N_n} Z_{\CC_I}=a_n \sum_{\gamma \in \Gamma_0(4) \backslash \SL_2(\Z)} U_\gamma\, \Psi_{0}+ {b_n\over 2^n}  \sum_{a,b=0,1} \Psi_{(2a,2b)},\eea
where for $n=1,2,3$,  $N_n=3,22,486$, $a_n=1,12,352$, and $b_n=1,16,544$.

Finally, we comment on the values of the $\alpha_I$. 
Our derivation proposes that all distinct states $|\Psi_{C_I}\rangle$ should enter with equal weights $\alpha_I=1$. This is correct as a bulk statement, but the weights of the individual CFTs in \eqref{duality} could vary due to accidental symmetries. 
In particular, a boundary state $\langle B|=\langle \xi,\bar \xi|$ could be invariant under a subgroup of invertible surface operators in the bulk, which we denote by $G_S$, 
\bea
\langle B|S=\langle B|, \qquad S\in G_S.
\eea
The group $G_S$ is the group of equivalences of topological boundary conditions in the sense that, provided $|\Psi_\CC\rangle$ and $|\Psi_\CC'\rangle$ are related by $G_S$,  $|\Psi_\CC\rangle$ and $|\Psi_\CC'\rangle$ both give rise to the same CFT. The action of $G_S$ can be uplifted to $\CC$, where it becomes the group of
equivalences of codes. 
A well-studied example is provided by $n$ copies of $k=2$ AB theory: in the particular case of the so-called rigid embedding $r=1$, $G_S$  consists of certain $T$-dualities which form a subgroup of the standard group of code equivalences  \cite{Dymarsky:2020qom}. In such a case, only inequivalent CFTs enter the sum \eqref{duality} with the weights
\bea
\alpha_I={|G_S|\over  | {\rm Aut}(\CC_I)  |} 
\eea
where $ {\rm Aut}(\CC_I)  \subset G_S$ is the group of symmetries of  any code $\CC$ in the equivalence class labeled by $I$. 

\section{Extensions and generalizations}
\label{sec:extensions}
\subsection{Ensembles of Narain theories }
The various examples of holographic Narain CFT ensembles recently discussed in the literature conform to the framework described in this work, with appropriate choices of the bulk TQFT.  The most straightforward 
 example is the one we refer to throughout  this work:  $n$ copies of the level-$k$ AB theory dual to a $\Z_k^n$ orbifold groupoid of Narain CFTs, as discussed in \cite{Aharony:2023zit,DHP}. The original case of $\U(1)$ gravity being dual to the average over all Narain theories \cite{Afkhami-Jeddi:2020ezh,Maloney:2020nni}  can be understood as the $k\rightarrow \infty$ limit of AB theory \cite{Aharony:2023zit}. Our construction thus answers two basic questions raised in the original work \cite{Maloney:2020nni}, about  the precise nature of the microscopic bulk description of $\U(1)$-gravity and about why  the bulk sum includes only handlebody topologies. Taking $k$ finite removes the problem with the divergent IR mode of the non-compact Abelian Chern-Simons theory. Then, certain observables, e.g.~the partition function average for $n\geq g+1$,  remain finite even in the $k\rightarrow \infty$ limit, although the finite-$k$ theory does not have such a limitation. 
 
The sum over topologies includes only handlebodies in the exceptional case of square-free $k$, when the transitive action of   $\rmO(n,n,\Z_k)$ implies that at any $g$ there is a unique state invariant under both $\rmO(n,n,\Z_k)$ and $\Sp(2g,\Z_k)$, which up to an overall normalization is given by the TQFT path integral summed only over handlebody topologies. This property survives the $k\rightarrow \infty$ limit, as is suggested by the transitive action of $\rmO(n,n,\R)$ on the Narain moduli space. 
In other words, the inclusion of only handlebodies in the context of U(1)-gravity is an accident. In principle, one should include all possible topologies, including wormholes,  but their contribution would merely  renormalize the overall coefficient of $\Psi_{\rm bulk}$, which is ignored in the approach of \cite{Afkhami-Jeddi:2020ezh,Maloney:2020nni}.

One might wonder if the $k\rightarrow \infty$ limit could depend on the way the limit is taken. In particular, if $k$ is not square-free, then for any finite $g$ additional topologies will contribute.   It would be interesting to see, starting with the Poincar\'e series of $\Psi_0^g$ for $g=2$ and reducing it to $g=1$, similarly to the calculation of section \ref{k=4}, whether these additional terms become negligibly small as $k$ increases.


Other examples in the literature also conform to the TQFT gravity framework of this paper. Among them
is the ensemble of rational Narain CFTs with $\U(1)_k$ chiral algebra \cite{Raeymaekers:2021ypf, Raeymaekers:2023ras}, which corresponds to taking $\Lambda_0$ to be the cubic lattice in $\R^{1,1}$ of size $\sqrt{k}$. The ensemble of rational Narain CFTs with $\SU(k)_1$ chiral algebra \cite{Meruliya:2021lul,Henriksson:2022dml} corresponds to $\Lambda_0$ being a direct sum of two root lattices of $\SU(k)$, etc. We consider these and other examples in detail in \cite{BDS}.

We expect that the ensembles consisting of various orbifolds of Narain theories \cite{Benjamin:2021wzr,Kames-King:2023fpa, Forste:2024zjt} would follow from the general construction with possibly additional degrees of freedom in the bulk. 
The construction of \cite{Ashwinkumar:2021kav,Ashwinkumar:2023ctt} calls for even but not necessarily self-dual codes, i.e.~non-maximal gaugings. 

We finally note that non-bosonic Abelian Chern-Simons theories should describe an ensemble of free fermionic CFTs. It would be interesting to develop this  in detail. 

\subsection{SymTFT construction}
The 2d CFTs  we have discussed up to now have been limited to Narain theories. We emphasize that all the relations discussed in section \ref{sec:bulk} are between states in the 3d Chern-Simons theory, while the Narain ensemble emerges from the choice of a particular basis in ${\cal H}^g$ corresponding to ``coherent'' boundary conditions. The case of Narain theories is indeed special because the level one AB Chern-Simons  theory  is  holographically dual to an individual  Narain theory, i.e.~the duality extends to include local operators and other physical quantities \cite{Gukov:2004id,Belov:2005ze,Benini:2022hzx,Ashwinkumar:2021kav,Aharony:2023zit}. In this correspondence, bulk gauge fields couple directly to scalar fields at the boundary, and hence to all eligible operators in the corresponding 2d theory. 

At the same time, following the paradigm of symTFT \cite{Freed:2022qnc}, one can couple a bulk topological field theory
to any boundary theory exhibiting appropriate  global symmetries. That would introduce a new basis of states in ${\cal H}^g$ in lieu
 of \eqref{basis}. As a standard example we consider the 2d Ising model, coupled to the $\Z_2$ ``toric code'' gauge theory given by \eqref{AB}  with $k=2$ \cite{Shao:2023gho}. Then the identity \eqref{TCP} becomes
 \bea
 \label{Ising}
 Z_{\rm Ising}(\tau)+Z_{\rm Ising}'(\tau)=\Psi_0(\tau)+\Psi_0(-1/\tau)+\Psi_0(-1/(\tau+1)),
 \eea
 where $Z_{\rm Ising}=\langle B|\Psi_{\CC_1}\rangle$ and $Z_{\rm Ising}'=\langle B|\Psi_{\CC_2}\rangle$ 
 are torus partition functions of the Ising model and its $\Z_2$ orbifold, which happen to be equal to each other by Kramers-Wannier duality
 \bea
 Z_{\rm Ising}=Z_{\rm Ising}'=|\chi_1(\tau)|^2+|\chi_\varepsilon(\tau)|^2+|\chi_\sigma(\tau)|^2.
 \eea 
 Here $\langle B|$ is the state in the 3d gauge theory created by coupling it to the Ising model on the boundary.
 The bulk wave-function can be written explicitly in terms of the Ising model characters as  \cite{Lin:2023uvm}
 \bea
 \label{IP}
 \langle B|\Psi_0\rangle =\Psi_0(\tau)=|\chi_1(\tau)|^2+|\chi_\varepsilon(\tau)|^2. 
 \eea 
 We note that \eqref{IP} does not represent a full holographic duality for the Ising theory, 
 since the gauge fields are not coupled to individual local operators of the Ising model. 
 
The relation \eqref{IP} extends to arbitrary $g$, with the Poincar\'e series being a sum over $\Gamma_{0}^g(2)\backslash \Sp(2g,\Z)$ applied to $\Psi_0^g$ on the RHS. We note that this is not the Poincar\'e series discussed in \cite{Castro:2011zq,Jian:2019ubz} because of the different seed, $\Psi_0^g \neq |\chi_1|^{2g}$. At the very least \eqref{IP} and its generalizations are useful mathematical identities, which can be translated as statements in Ising TQFT. 

\subsection{Higher dimensions}
The general idea behind \eqref{bulk} is not specific to 3d and can be applied to higher-dimensional theories. One concrete example is the 5d $\U(1)^g$ BF theory dual to $\U(1)^g$ Maxwell theory in 4d \cite{Witten:1995gf,Verlinde:1995mz}. Starting from the level-$k$ theory in 5d, with the boundary geometry being the connected sum of  $n$ copies of $\Sphere^2\times \Sphere^2$, one finds the same mathematical structure as in the case of level-$k$ AB theory in 3d. The main difference is in interpretation of the corresponding groups, which exchange their roles.  The mapping class group $\mathscr{M}$ of the 4-geometry is $\rmO(n,n,\Z_k)$, while different gaugings of Maxwell theory \cite{Choi:2021kmx} are permuted by $\Sp(2g,\Z_k)$. Demanding that all 4d theories enter the ensemble with equal weights, and taking $k=4$ as an example would lead to equal coefficients $a_n=b_n$ in \eqref{k4P}, forcing non-trivial coefficients $\alpha_I$ (which now weight different topologies) in the LHS. This example will be discussed in detail elsewhere.


\subsection{Non-Abelian TQFTs}
Finally, we discuss the case of a general 3d TQFT. The ensemble of boundary CFTs will be defined as arising from all possible topological boundary conditions in the bulk theory. The corresponding states, denoted by $|\Psi_I\rangle$, can be defined at any genus $g$ and are modular invariant. In addition, at any given finite $g$, there could be other modular invariant states linearly independent from $|\Psi_I\rangle$. In terms of the boundary theory, these additional states are non-physical modular invariant combinations of conformal blocks, which are well known for $g=1$ in the context 
of minimal models or $\SU(2)_k$ WZW theory \cite{cappelli1987modular,cappelli1987ade}. We expect that as $g$ increases the number of non-physical modular invariants will decrease to zero, which is a rephrasing of the expectation that the conformal modular bootstrap at arbitrarily high genus is powerful enough to allow only well-defined CFTs as solutions. An explicit example of this behavior is provided by  $\SU(2)_{k} \times{\SU(2)}_{-k}$ Chern-Simons theory \cite{Korinman_2019}. 

In the $g\rightarrow \infty$ limit we therefore expect that $|\Psi_I\rangle$ are the only states  invariant under the modular group; moreover their normalized versions $|\psi_I\rangle=|\Psi_I\rangle^g/|\Psi_I^g|$ will be mutually orthonormal. We thus arrive at the same identity in the bulk \eqref{bulk}, with the finite-$g$ version defined, at least in principle, by genus reduction. There are several important points which make the non-Abelian case significantly more intricate. First, in general the representation of the mapping class group will have infinite order $|\mathscr{M}_g|$, rendering the corresponding sum in \eqref{bulk} ill-defined \cite{Jian:2019ubz}.  Also there is no simple formula for the norm $|\Psi_I^g|$ at finite $g$, where we define the corresponding state to include exactly one vacuum character ${}^g\langle 0|\Psi^g_I\rangle=1$. This means that even though the weights are the same in \eqref{bulk}, they may become different once the normalization of $|\Psi_I^g\rangle$ is taken into account. Subsequent genus reduction via $\Phi$ will not change  the weights, suggesting 
\bea
{\alpha_I \over \alpha_J} =\lim_{g\rightarrow \infty} {|\Psi_J^g|^2\over |\Psi_I^g|^2}.
\eea
Finally, in the non-Abelian case the Hilbert space ${\cal H}^g$ is non-factorizable. We expect the genus reduction through  ${}^{\tilde g}\langle 0|U_\gamma|0\rangle^{g+\tilde{g}}$ in the limit $\tilde{g}\rightarrow \infty$ will include all possible wormhole topologies, but evaluation of the corresponding expression in any given theory could be hard. 

We note that in the non-Abelian case genus reduction through $\Phi$ and partial trace 
will lead to different results. While $\Phi$ preserves the weights, taking the partial  trace of $\rho_{g+\tilde{g}}$ may change relative weights in the reduced density matrix $\rho_g$. This would  lead  to genus-dependent entropy $S_g=-{\rm Tr}_g\, \rho_g \ln \rho_g$ due to holographic  entanglement, making an interesting connection with the  ``entanglement  from topology'' literature  \cite{Salton:2016qpp,balasubramanian2017multi,balasubramanian2018entanglement,melnikov2019topological,PhysRevD.107.126005,Melnikov:2023wwc}.

\section{Discussion}
\label{sec:discussion}
In this paper we have advocated a holographic duality between an ensemble of  2d CFTs
on a genus-$g$ surface $\Sigma_g$ and 
``TQFT gravity'' in the bulk, with the TQFT path integral summed over topologically distinct  3-manifolds $M$ with boundary $\Sigma$.\footnote{{The duality extends to corralators of local operators, although we do not discuss this here.}} The boundary ensemble consists of all  maximal gaugings (states defined by topological boundary conditions) of the original TQFT; thus, the duality can be formulated entirely in the bulk. We have outlined a general derivation of this duality, which does not rely on evaluating the boundary average or the bulk ``Hartle-Hawking'' wavefunction explicitly, e.g.~by using variants of the Siegel-Weil formula. Our approach has been to consider the limit of large genus, in which the duality is established by writing the projector onto the modular invariant subspace of the TQFT Hilbert space in two different ways \eqref{bulk}. One of these representations gives rise to a gravitational sum over bulk topologies. 
 The finite-$g$ version is then obtained via genus reduction. The corresponding sum in the bulk, at least in principle, includes all possible topologies,\footnote{If $\Sigma$ is orientable, only orientable ${M}$ appear in the sum.} including singular topologies resulting from genus reduction.  

In the case of a bosonic, non-anomalous 3d Abelian Chern-Simons theory with compact gauge group and proper boundary conditions at $\Sigma=\partial {M}$, the boundary ensemble is a discrete set of Narain ``code CFTs,''  discussed in \cite{Dymarsky:2020qom,yahagi2022narain,angelinos2022optimal,Aharony:2023zit,Kawabata:2023iss}.  At each $g$, the sum over 3d topologies  reduces to a finite sum over the set of equivalence classes, with potentially different weights.   
In each class one can choose a representative which is either a handlebody  or a  handlebody with  insertions of particular sets of Wilson lines. These equivalence classes, and  the TQFT path integrals on these topologies,  are  parametrized by symplectic codes.  The appearance of codes on both sides of the duality \eqref{duality}
suggests an interpretation  in terms of quantum information \cite{BDS}.

In the original studies of the ensemble of all Narain CFTs \cite{Afkhami-Jeddi:2020ezh,Maloney:2020nni}, the prescription to include only the contributions of  handlebody topologies was left unexplained.  We have clarified previously \cite{Aharony:2023zit} that the bulk TQFT is actually the compact $U(1)^{c}\times U(1)^c$ Chern-Simons theory, in the limit of taking level $k$ to infinity. This theory should actually be summed over {\it all} topologies, even including singular ones. For finite $k$ the theory can only distinguish finitely many equivalence classes of topologies, and in the $k\to \infty$ limit  the equivalence classes are represented by all possible handlebodies.  
Thus summing over  all topologies versus just handlebodies will  affect  only the overall normalization of the bulk wavefunction, which is automatically  fixed by requiring a unique vacuum state. 

Our proposal provides a recipe, at least in principle,  for extending the duality to non-Abelian TQFTs, including 
cases like, e.g.,~$\SU(2)_k \times {\SU(2)}_{-k}$   \cite{Meruliya:2021utr}  or Virasoro TQFT \cite{maloney2010quantum,keller2015poincare}, which previously suffered from  non-unitarity, due to  presence of unphysical invariants or negative weights in the sum \eqref{duality},  or  non-positive density of states. The case of RCFTs is conceptually simpler because the  TQFT Hilbert space ${\cal H}^g$ and the spectrum of Wilson line operators are finite.  Thus, genus reduction ${}^{\tilde g}\langle 0|U_\gamma|0\rangle^{g+\tilde{g}}$ for any mapping class group element $\gamma$ is well-defined. By starting at (infinitely) large genus and then performing a genus reduction, we should end up with only physical theories in the boundary ensemble, positive weights $\alpha_I$, and hence a positive density of states. 

This recipe should also extend  to the Virasoro TQFT \cite{maloney2010quantum}, although this case is more challenging because  the spectrum of  line operators is continuous. One can envision starting from the Virasoro vacuum character on a genus-2 surface $\Sigma$, performing a modular transformation $\gamma$ and then performing the genus reduction to $g=1$. This calculation would be schematically similar to the one performed in section \ref{k=4}. In the general case, the resulting geometry would be as shown  in the right panel of Fig.\ref{fig:wormhole} -- a solid torus with  a line defect wrapping a non-shrinkable cycle. One can think of it as thermal AdS${}_3$ with a  static point-like object whose worldline forms a line defect. In certain cases the defect will be entirely absent or will be a BTZ or a dyonic black hole, resulting in a smooth handlebody geometry.  The latter two cases are qualitatively similar to the case of $k=2$ theory discussed in section \ref{k=4}. In other cases the geometry will be singular. This calculation would not be sufficient to identify the correct seed of the Poincar\'e series, but it would outline the set of singular geometries with line defects contributing to the  bulk sum. It would be interesting to compare and match these geometries with the conical defects and other non-handlebody configurations  proposed to cure the partition function of pure gravity \cite{benjamin2020pure,Maxfield:2020ale,PhysRevLett.132.041602}. Another related issue which our approach can potentially resolve is the discrepancy between the Virasoro TQFT amplitudes on non-hyperbolic geometries with those of pure 3d gravity. Taking the trumpet amplitude as an example \cite{Cotler:2020ugk}, the conventional  Virasoro TQFT calculation yields an incorrect result (see \cite{Jafferis:2024jkb} for a recent discussion). By contrast our approach would start with a sum over 3d topologies ending on a genus-2 Riemann surface, and then deform it into two disjoint tori. As we have seen in section  \ref{k=4}, the resulting trumpet geometry would include surface operator insertions as in   \cite{Jafferis:2024jkb}, conceivably leading to the correct result.

\acknowledgments
We thank Scott Collier, Tom Hartman, Shu-Heng Shao and especially Ahmed Barbar for  discussions.
AD is thankful to Institute Pascal for hospitality, where this work was completed. 
AD  is supported by the NSF grant PHY 2310426.


\appendix
\section{Abelian Chern-Simons theory}
\label{sec:CS}
As is discussed in the main text, the Wilson line operator around cycle $\upgamma$ is defined to be 
\bea
\label{WL}
W_c(\upgamma)={\rm exp} \left \{ i \oint_\upgamma c_\alpha A^\alpha \right\}, \qquad c\in \D\equiv \Lambda_0^*/\Lambda_0,
\eea
where the charges are vectors $c\in \Lambda_0^*$ modulo identifications $c\sim c+v$ for any $v\in \Lambda_0$. The scalar product $K_{ij}$ on $\Lambda_0$ defines the scalar product $\eta$ on $\Lambda_0^*$ and $\D$ (we call it $\eta$ rather than $K^{-1}$ to match the recent Narain code CFT literature). For the case of $n$ copies of level $k$ AB theory the scalar product can be written explicitly as
\bea
\eta={1\over k}\left(
\begin{array}{cc}
0 & 1_n\\
1_n & 0
\end{array}
\right).
\eea

The intersection numbers of one-cycles $\upgamma_i$ on a Riemann surface $\Sigma$ of genus $g$ are given by the symplectic form $\omega(\upgamma_i,\upgamma_j)$ , which can be brought to canonical form  
\bea
\label{sp}
\omega=\left(
\begin{array}{cc}
0 & -1_g\\
1_g & 0
\end{array}
\right),
\eea
by choosing first $g$ cycles to be $a$-cycles and the last $g$ to be $b$-cycles. 

Wilson lines operators \eqref{WL} satisfy the following commutation relation 
\bea
W_c(\upgamma) W_{c'}(\upgamma') =W_{c'}(\upgamma') W_c(\upgamma) e^{2\pi i \eta(c,c')\omega(\upgamma,\upgamma')}.
\eea
Wilson lines wrapping $a$- and $b$-cycles generate the group ${\cal W}$ of all Wilson lines,  called the magnetic translation group in \cite{Belov:2005ze}. Up to an overall phase, the most general group element is
\bea
\label{WLg}
W[c_1,\dots,c_g,c_1',\dots,c_g']=W_{c_1}(a_1)\dots W_{c_g}(a_g) W_{c_1'}(b_1)\dots W_{c_g'}(b_g)
\eea
In the case of AB theory this is the ``Pauli'' group of standard shift and clock operators acting on $2ng$ qudits of dimension $k$ \cite{DHP}. 

The group $\cal W$ has automorphisms of various kinds.  One subgroup of ${\rm Aut}({\cal W})$ corresponds to the automorphisms of $\D$ that preserve $\eta$. This is the group $\rmO(n,n,\Z_k)$ in the case of level-$k$ AB theory. Another subgroup of ${\rm Aut}({\cal W})$  is the modular group $\Sp(2g,\Z)$, which reduces to $\Sp(2g,\Z_k)$ in the AB case. These two subgroups form reductive dual pair inside the full group of automorphisms $\Sp(4ng,\Z_k)$. (Strictly speaking this group acts on the labels $(c_1,\dots,c_g')$ in \eqref{WLg}. The action on Wilson line operators may include a phase.)

A basis for ${\cal H}^g$ -- the Hilbert space of Chern-Simons wave-functions on $\Sigma$ -- is given by the path integral on a handlebody with shrinkable $a$-cycles and Wilson lines wrapping $b$-cycles,
\bea
|\Psi_{c_1\dots c_g}\rangle\equiv |c_1\dots c_g\rangle=\int {\mathcal D} A\,  e^{i S}\, \prod_{i=1}^g W_{c_i}(b_i).
\eea
This is the same as $\prod_{i=1}^g W_{c_i}(b_i)$ acting on $|0\rangle^g \equiv |\Psi_{0\dots 0}\rangle$ -- the path integral on the handlebody without insertions.  
In this basis the action of  an individual Wilson line is given by 
\bea
&&W_c(a_\ell)|c_1\dots c_g\rangle=e^{2\pi i\, \eta(c,c_\ell)} |c_1\dots c_g\rangle,\\
&&W_c(b_\ell)|c_1\dots c_\ell \dots c_g\rangle=|c_1\dots c_\ell+c\dots  c_g\rangle.
\eea
It is straightforward to see that $W_c(a_\ell)$ for any $c\in \D$ and $\ell$ act trivially on $|0\rangle$ which is consistent with  its definition as a path integral on a topology where all $a_i$ are shrinkable. 

Wavefunctions $\Psi_{c_1\dots c_g}(\Omega,\xi,\bar\xi)$ can be written explicitly in terms of the boundary conditions for the bulk  gauge fields introduced in \cite{Aharony:2023zit}
\bea
\Psi_c(\tau,\xi,\bar \xi)={1 \over |\eta(\tau)|^{2n}} \sum_{v\in \Lambda_c} e^{i \pi \tau p_L^2-i \pi \bar \tau p_R^2+2\pi i (p_L \cdot \xi -p_R\cdot \bar \xi) +{\pi \over 2\tau_2}(\xi^2+\bar\xi^2)},
\eea
where $\Lambda_c$ is defined in \eqref{lambdac} and $v\equiv (p_L,p_R)$  is in the basis where $\eta$ is diagonal. 
An alternative approach is to introduce a Euclidean structure on $\Lambda_0$ by adding in a Maxwell term in the bulk, which is irrelevant and decouples in the low energy limit \cite{Gukov:2004id,Belov:2005ze}. The same expression emerges from the bottom-up code construction. Its generalization to higher genus is straightforward \cite{Henriksson:2021qkt,Henriksson:2022dnu,Belov:2005ze}
\bea
\Psi_{c_1\dots c_g}(\Omega,\xi,\bar \xi) ={1\over  \Phi(\Omega)} 
\sum_{v^i\in \Lambda_{c_i}} 
e^{i\pi (p_L^i \Omega_{ij}p_L^j - p_R^i\bar\Omega_{ij}p_R^j)
+2\pi i (p^i_L \cdot \xi^i-p^i_R \cdot \bar \xi^i )+ \frac{\pi}{2} \, (\Im(\Omega))^{-1}_{ij}( \xi^T_i \xi_j +\bar \xi^T_i  \bar \xi_j) }.\nonumber\\
\eea
Here $\Omega_{ij}$ is the modular parameter of $\Sigma$ and $\Phi(\Omega)$ is due to  small oscillations of bulk gauge fields.

\section{Contributing topologies and genus reduction}
\label{genusreduction}
The set of inequivalent handlebody topologies $\cal M$ ending on a Riemann surface $\Sigma=\partial \cal M$ of genus $g$ can be parametrized by maximal Lagrangian sublattices in $H_1(\Sigma,\Z)$ equipped with the symplectic product \eqref{sp},
as discussed e.g.~in \cite{Maloney:2020nni}.

We would like to  generalize this  description to include other topologies, including singular ones, such as those discussed in section \ref{k=4}. In what follows,  we focus for simplicity on the case of ($n$ copies of) level $k$ AB theory. An observation  of \cite{Barbar:2023ncl} was that Lagrangian lattices can be understood as a particular case of  symplectic codes, which we define as all possible Lagrangian subgroups $\cal L$ of $H_1(\Sigma,\Z_k)$, self-dual with respect to  symplectic product \eqref{sp}. It is easy to see that any Lagrangian lattice in $H_1(\Sigma,\Z)$ gives rise to such a code  $\cal L$. 
Simplest examples is given by ${\cal L}_0=(a_1,\dots,a_g,0,\dots,0)$ with arbitrary $a_i \in \Z_k$, which corresponds to the particular handlebody topology with shrinkable $a$ cycles.  To establish this identification we first define  a subgroup 
\bea
\label{qfroml}
Q_{\cal L}={\cal L} \otimes \D  \subset \underbrace{\D \oplus \dots  \oplus \D}_{2g},
\eea
where we keep in mind that $\D=\Z_k^{2n}$ and multiplication in  $Q_{\cal L}$ is defined mod $k$. For example,  in case of ${\cal L}_0$, this group includes all elements of the form 
\bea
\label{L0}
(c_1,\dots, c_g,0, \dots,0)\in Q_{{\cal L}_0},\qquad c_i\in \D.
\eea
In full generality the group $Q_{\cal L}$ is even self-dual with respect to $\eta \otimes w$ and defines  a maximal commutative subgroup  ${\cal W}_{\cal L}$ inside the group of Wilson line operators
\bea
W[c_1\dots c_g,c_1' \dots c_g'] \in  {\cal W}_{\cal L}\subset {\cal W},\qquad (c_1\dotsc_g,c_1'\dots c_g')\in Q_{{\cal L}}.
\nonumber
\eea
Since the group ${\cal W}_{\cal L}$ \,is a maximal Abelian subgroup of $\cal W$, it defines a stablizer state $|\Psi_{\cal L}\rangle$ satisfying
\bea
W |\Psi_{\cal L}\rangle=|\Psi_{\cal L}\rangle\qquad
\forall \,W\in {\cal W}_{\cal L}\, .
\eea
In other words ${\cal W}_{\cal L}$ is a quantum stablizer code defined in terms of the classical ``symplectic code'' $\cal L$. This picture is further developed in \cite{Barbar:2023ncl}.
Going back to \eqref{L0} it is clear that  $|\Psi_{{\cal L}_0}\rangle=|0\rangle^g$, which is indeed a stablizer state \cite{Salton:2016qpp}.

When $k$ is square-free, all symplectic codes $\cal L$ and all corresponding states are obtained by modular transformations $\Sp(2g,\Z_k)$ acting on ${\cal L}_0$ and $|0\rangle^g$ correspondingly. Thus the only  topologies contributing to \eqref{duality} or \eqref{PS} are the handlebodies. 
But when $k$ is not square-free, there are additional classes of $\cal L$. For example, for $g=1$ and $k=p^2$ with prime $p$ there is an additional $\cal L$ not in the $\SL(2,\Z_k)$ orbit of ${\cal L}_0$: 
\bea
{\cal L}_p =\{(a,b)\, |\, a,b\in \Z_k,\,  a,b \equiv0~ {\rm mod}\,  p\}.
\eea
The corresponding state is, c.f.~\eqref{topologies},
\bea
|\Psi_{{\cal L}_p}\rangle={1\over p^{n}} \sum_{a_i,b_i=0}^{p-1} |\Psi_{(pa_1, pb_1\dots)}\rangle.
\eea
In general, for $g=1$ the symplectic self-dual codes ${\cal L}$ are in one-to-one correspondence with terms in the definition of the Hecke operator $T_k$. 

To finish the discussion of topologies, we need to show that the states $|\Psi_{{\cal L}}\rangle$ for all possible $\cal L$ are the only states appearing from the genus reduction. To this end we sketch the calculation of ${}^{\tilde g}\langle 0|U_\gamma|0\rangle^{g+\tilde{g}}$. Schematically, it goes as follows. The state $|0\rangle^{g+\tilde{g}}$ is defined by the Lagrangian group $Q_{{\cal L}_0}\subset \D^{2g+2\tilde{g}}$. Acting on it with $U_\gamma$ turns it into a state defined in terms of some other Lagrangian group 
\bea
(c_1\dots c_{g+\tilde{g}},c'_1,\dots c'_{g+\tilde{g}})\in Q_{\cal L'},
\eea
where $(c_1\dots c'_{g+\tilde{g}})$ is related to $(\tilde{c}_1\dots \tilde{c}_{g+\tilde{g}},0\dots 0)$ with arbitrary $\tilde{c}_i$ by the transformation $\gamma\in \Sp(2g+2\tilde{g},\Z_k)$. Next, taking the scalar product with ${}^{\tilde{g}}\langle 0|$ is equivalent to taking the subset of elements of $Q_{\cal L'}$ with $c_1=\dots =c_{\tilde{g}}=0$ and omitting them and $c_1',\dots,c_{\tilde g}'$,
\bea
(c_1\dots c_{g+\tilde{g}},c'_1,\dots c'_{g+\tilde{g}}) \rightarrow (c_{\tilde{g}+1}\dots c_{g+\tilde{g}},c'_{\tilde{g}+1},\dots c'_{g+\tilde{g}}).
\eea 
It is easy to see that this map defines a Lagrangian subgroup $Q \subset \D^{2g}$. A less obvious question is to see that this subgroup is defined by \eqref{qfroml} by some appropriate $\cal L$. The state ${}^{\tilde g}\langle 0|U_\gamma|0\rangle^{g+\tilde{g}}$ is then proportional to $|\Psi_{\cal L}\rangle$. 

There are usually many different $\gamma\in \Sp(2g+2\tilde{g},\Z_k)$ which will result in the  same $Q=Q_{\cal L}$ with the same ${\cal L}(\gamma)$, but the proportionality coefficient in 
\bea
{}^{\tilde g}\langle 0|U_\gamma|0\rangle^{g+\tilde{g}} \propto |\Psi_{{\cal L}(\gamma)}\rangle, 
\eea
which depends on how many pre-images each element in $Q_{\cal L}$ has inside $Q_{\cal L'}$, will normally depend on the equivalence class of 
$\gamma$ inside $\Sp(2g,\Z_k)$. Thus,  identifying all states potentially contributing to the seed $\Psi_{\rm seed}$ of Poincare series, i.e.~the equivalence classes of topologies contributing to the bulk sum, is straightforward, but evaluating corresponding coefficients starting from high genus $g+\tilde{g}$ is less trivial. 

We finally address the question of how  many free parameters can appear in the duality identity  \eqref{PS}, assuming one requires invariance under both $\rmO(n,n,\Z_k)$ and $\Sp(2g,\Z_k)$ groups. As was discussed in section \ref{k=4}, codes $\CC$ and symplectic codes $\cal L$ are abelian groups,  e.g.~$\Z_k^{J-m} \Z_p^{2m}$ for $J=n$ and $J=g$ respectively when $k=p^2$. Clearly, the linear action of  $\rmO(n,n,\Z_k)$ and $\Sp(2g,\Z_k)$ on $\CC$ and $\cal L$ can not change their group structure, but otherwise acts transitively. Thus codes fall into $n+1$ and $g+1$ distinct orbits of $\rmO(n,n,\Z_k)$  and  $\Sp(2g,\Z_k)$, respectively.  Since the states $|\Psi_\CC\rangle^g$ are invariant under $\Sp(2g,\Z_k)$ and $|\Psi_{\cal L}\rangle$ are invariant under $\rmO(n,n,\Z_k)$ we readily conclude there are $n+1$ distinct coefficients $\alpha_I$ controlling the LHS of  \eqref{PS} and $g+1$ coefficients weighting different topologies on the RHS of \eqref{PS}. This is a bit puzzling, since the number of free parameters on the LHS and RHS of \eqref{PS} should be the same. The resolution comes from the fact that the states $|\Psi_\CC\rangle^g$ as well as $|\Psi_{\cal L}\rangle$ can be linearly dependent. To illustrate, consider the case of $g=1$. There are $k+p+1$ distinct orbits of $\rmO(n,n,\Z_k)$ contained in $\D$, corresponding to the $k+p+1$ codes $\cal L$, which form $2$  orbits with respect to $\SL(2,\Z_k)$. There are many codes $\CC$;\footnote{The exact counting is non-trivial, see \cite{gaborit1996mass} for a  related problem.} each code gives rise to a modular-invariant $|\Psi_\CC\rangle$. Summing $|\Psi_\CC\rangle$ over an $\rmO(n,n,\Z_k)$ orbit of $\CC$  will give a state invariant under both groups. But all of these states are linear superpositions of just two states; hence for $n>1$ they are linearly dependent. Thus only $2$ specific combinations made out of the $n+1$  coefficients $\alpha_I$ appear on the LHS of \eqref{PS}.

\bibliographystyle{JHEP}
\bibliography{bulk.bib}

\providecommand{\href}[2]{#2}\begingroup\raggedright\begin{thebibliography}{10}

\bibitem{Castro:2011zq}
A.~Castro, M.R.~Gaberdiel, T.~Hartman, A.~Maloney and R.~Volpato, \emph{{The
  Gravity Dual of the Ising Model}},
  \href{https://doi.org/10.1103/PhysRevD.85.024032}{\emph{Phys. Rev. D}
  {\bfseries 85} (2012) 024032}
  [\href{https://arxiv.org/abs/1111.1987}{{\ttfamily 1111.1987}}].

\bibitem{Jian:2019ubz}
C.-M.~Jian, A.W.W.~Ludwig, Z.-X.~Luo, H.-Y.~Sun and Z.~Wang,
  \emph{{Establishing strongly-coupled 3D AdS quantum gravity with Ising dual
  using all-genus partition functions}},
  \href{https://doi.org/10.1007/JHEP10(2020)129}{\emph{JHEP} {\bfseries 10}
  (2020) 129} [\href{https://arxiv.org/abs/1907.06656}{{\ttfamily
  1907.06656}}].

\bibitem{Romaidis:2023zpx}
I.~Romaidis and I.~Runkel, \emph{{CFT correlators and mapping class group
  averages}},  \href{https://arxiv.org/abs/2309.14000}{{\ttfamily 2309.14000}}.

\bibitem{Afkhami-Jeddi:2020ezh}
N.~Afkhami-Jeddi, H.~Cohn, T.~Hartman and A.~Tajdini, \emph{{Free partition
  functions and an averaged holographic duality}},
  \href{https://doi.org/10.1007/JHEP01(2021)130}{\emph{JHEP} {\bfseries 01}
  (2021) 130} [\href{https://arxiv.org/abs/2006.04839}{{\ttfamily
  2006.04839}}].

\bibitem{Maloney:2020nni}
A.~Maloney and E.~Witten, \emph{{Averaging over Narain moduli space}},
  \href{https://doi.org/10.1007/JHEP10(2020)187}{\emph{JHEP} {\bfseries 10}
  (2020) 187} [\href{https://arxiv.org/abs/2006.04855}{{\ttfamily
  2006.04855}}].

\bibitem{Datta:2021ftn}
S.~Datta, S.~Duary, P.~Kraus, P.~Maity and A.~Maloney, \emph{{Adding flavor to
  the Narain ensemble}},
  \href{https://doi.org/10.1007/JHEP05(2022)090}{\emph{JHEP} {\bfseries 05}
  (2022) 090} [\href{https://arxiv.org/abs/2102.12509}{{\ttfamily
  2102.12509}}].

\bibitem{cotler2020ads3}
J.~Cotler and K.~Jensen, \emph{Ads3 wormholes from a modular bootstrap},
  {\emph{Journal of High Energy Physics} {\bfseries 2020} (2020) 1}.

\bibitem{dymarsky2021comments}
A.~Dymarsky and A.~Shapere, \emph{Comments on the holographic description of
  narain theories}, {\emph{Journal of High Energy Physics} {\bfseries 2021}
  (2021) 1}.

\bibitem{Benjamin:2021wzr}
N.~Benjamin, C.A.~Keller, H.~Ooguri and I.G.~Zadeh, \emph{{Narain to Narnia}},
  \href{https://doi.org/10.1007/s00220-021-04211-x}{\emph{Commun. Math. Phys.}
  {\bfseries 390} (2022) 425}
  [\href{https://arxiv.org/abs/2103.15826}{{\ttfamily 2103.15826}}].

\bibitem{collier2022wormholes}
S.~Collier and A.~Maloney, \emph{Wormholes and spectral statistics in the
  narain ensemble}, {\emph{Journal of High Energy Physics} {\bfseries 2022}
  (2022) 1}.

\bibitem{Ashwinkumar:2021kav}
M.~Ashwinkumar, M.~Dodelson, A.~Kidambi, J.M.~Leedom and M.~Yamazaki,
  \emph{{Chern-Simons invariants from ensemble averages}},
  \href{https://doi.org/10.1007/JHEP08(2021)044}{\emph{JHEP} {\bfseries 08}
  (2021) 044} [\href{https://arxiv.org/abs/2104.14710}{{\ttfamily
  2104.14710}}].

\bibitem{Ashwinkumar:2023jtz}
M.~Ashwinkumar, J.M.~Leedom and M.~Yamazaki, \emph{{Duality Origami: Emergent
  Ensemble Symmetries in Holography and Swampland}},
  \href{https://arxiv.org/abs/2305.10224}{{\ttfamily 2305.10224}}.

\bibitem{Ashwinkumar:2023ctt}
M.~Ashwinkumar, A.~Kidambi, J.M.~Leedom and M.~Yamazaki, \emph{{Generalized
  Narain Theories Decoded: Discussions on Eisenstein series, Characteristics,
  Orbifolds, Discriminants and Ensembles in any Dimension}},
  \href{https://arxiv.org/abs/2311.00699}{{\ttfamily 2311.00699}}.

\bibitem{Kames-King:2023fpa}
J.~Kames-King, A.~Kanargias, B.~Knighton and M.~Usatyuk, \emph{{The Lion, the
  Witch, and the Wormhole: Ensemble averaging the symmetric product orbifold}},
   \href{https://arxiv.org/abs/2306.07321}{{\ttfamily 2306.07321}}.

\bibitem{Aharony:2023zit}
O.~Aharony, A.~Dymarsky and A.D.~Shapere, \emph{{Holographic description of
  Narain CFTs and their code-based ensembles}},
  \href{https://arxiv.org/abs/2310.06012}{{\ttfamily 2310.06012}}.

\bibitem{Forste:2024zjt}
S.~Forste, H.~Jockers, J.~Kames-King, A.~Kanargias and I.G.~Zadeh,
  \emph{{Ensemble Averages of $\mathbb{Z}_2$ Orbifold Classes of Narain CFTs}},
   \href{https://arxiv.org/abs/2403.02976}{{\ttfamily 2403.02976}}.

\bibitem{Meruliya:2021utr}
V.~Meruliya, S.~Mukhi and P.~Singh, \emph{{Poincar\'e Series, 3d Gravity and
  Averages of Rational CFT}},
  \href{https://doi.org/10.1007/JHEP04(2021)267}{\emph{JHEP} {\bfseries 04}
  (2021) 267} [\href{https://arxiv.org/abs/2102.03136}{{\ttfamily
  2102.03136}}].

\bibitem{maloney2010quantum}
A.~Maloney and E.~Witten, \emph{Quantum gravity partition functions in three
  dimensions}, {\emph{Journal of High Energy Physics} {\bfseries 2010} (2010)
  1}.

\bibitem{giombi2008one}
S.~Giombi, A.~Maloney and X.~Yin, \emph{One-loop partition functions of 3d
  gravity}, {\emph{Journal of High Energy Physics} {\bfseries 2008} (2008)
  007}.

\bibitem{keller2015poincare}
C.A.~Keller and A.~Maloney, \emph{Poincar{\'e} series, 3d gravity and cft
  spectroscopy}, {\emph{Journal of High Energy Physics} {\bfseries 2015} (2015)
  1}.

\bibitem{mikhaylov2018teichmuller}
V.~Mikhaylov, \emph{Teichm{\"u}ller tqft vs. chern-simons theory},
  {\emph{Journal of High Energy Physics} {\bfseries 2018} (2018) 1}.

\bibitem{Chandra:2022bqq}
J.~Chandra, S.~Collier, T.~Hartman and A.~Maloney, \emph{{Semiclassical 3D
  gravity as an average of large-c CFTs}},
  \href{https://doi.org/10.1007/JHEP12(2022)069}{\emph{JHEP} {\bfseries 12}
  (2022) 069} [\href{https://arxiv.org/abs/2203.06511}{{\ttfamily
  2203.06511}}].

\bibitem{Collier:2023cyw}
S.~Collier, L.~Eberhardt, B.~Muehlmann and V.A.~Rodriguez, \emph{{The Virasoro
  minimal string}},
  \href{https://doi.org/10.21468/SciPostPhys.16.2.057}{\emph{SciPost Phys.}
  {\bfseries 16} (2024) 057}
  [\href{https://arxiv.org/abs/2309.10846}{{\ttfamily 2309.10846}}].

\bibitem{collier2023solving}
S.~Collier, L.~Eberhardt and M.~Zhang, \emph{Solving 3d gravity with virasoro
  tqft}, {\emph{SciPost Physics} {\bfseries 15} (2023) 151}.

\bibitem{collier20243d}
S.~Collier, L.~Eberhardt and M.~Zhang, \emph{3d gravity from virasoro tqft:
  Holography, wormholes and knots}, {\emph{arXiv preprint arXiv:2401.13900}
  (2024) }.

\bibitem{Chen:2024unp}
L.~Chen, L.-Y.~Hung, Y.~Jiang and B.-X.~Lao, \emph{{Quantum 2D Liouville
  Path-Integral Is a Sum over Geometries in AdS$_3$ Einstein Gravity}},
  \href{https://arxiv.org/abs/2403.03179}{{\ttfamily 2403.03179}}.

\bibitem{Hung:2024gma}
L.-Y.~Hung and Y.~Jiang, \emph{{Building up quantum spacetimes with BCFT
  Legos}},  \href{https://arxiv.org/abs/2404.00877}{{\ttfamily 2404.00877}}.

\bibitem{deBoer:2024kat}
J.~de~Boer, D.~Liska and B.~Post, \emph{{Multiboundary wormholes and OPE
  statistics}},  \href{https://arxiv.org/abs/2405.13111}{{\ttfamily
  2405.13111}}.

\bibitem{Barbar:2023ncl}
A.~Barbar, A.~Dymarsky and A.D.~Shapere, \emph{{Global Symmetries, Code
  Ensembles, and Sums Over Geometries}},
  \href{https://arxiv.org/abs/2310.13044}{{\ttfamily 2310.13044}}.

\bibitem{Belov:2005ze}
D.~Belov and G.W.~Moore, \emph{{Classification of Abelian spin Chern-Simons
  theories}},  \href{https://arxiv.org/abs/hep-th/0505235}{{\ttfamily
  hep-th/0505235}}.

\bibitem{RUNGE1996175}
B.~Runge, \emph{Codes and siegel modular forms},
  \href{https://doi.org/https://doi.org/10.1016/0012-365X(94)00271-J}{\emph{Discrete
  Mathematics} {\bfseries 148} (1996) 175}.

\bibitem{Kraus:2006nb}
P.~Kraus and F.~Larsen, \emph{{Partition functions and elliptic genera from
  supergravity}},
  \href{https://doi.org/10.1088/1126-6708/2007/01/002}{\emph{JHEP} {\bfseries
  01} (2007) 002} [\href{https://arxiv.org/abs/hep-th/0607138}{{\ttfamily
  hep-th/0607138}}].

\bibitem{angelinos2022optimal}
N.~Angelinos, D.~Chakraborty and A.~Dymarsky, \emph{Optimal narain cfts from
  codes}, {\emph{Journal of High Energy Physics} {\bfseries 2022} (2022) 1}.

\bibitem{Dymarsky:2020qom}
A.~Dymarsky and A.~Shapere, \emph{{Quantum stabilizer codes, lattices, and
  CFTs}}, \href{https://doi.org/10.1007/JHEP03(2021)160}{\emph{JHEP} {\bfseries
  21} (2020) 160} [\href{https://arxiv.org/abs/2009.01244}{{\ttfamily
  2009.01244}}].

\bibitem{Benini:2022hzx}
F.~Benini, C.~Copetti and L.~Di~Pietro, \emph{{Factorization and global
  symmetries in holography}},
  \href{https://doi.org/10.21468/SciPostPhys.14.2.019}{\emph{SciPost Phys.}
  {\bfseries 14} (2023) 019}
  [\href{https://arxiv.org/abs/2203.09537}{{\ttfamily 2203.09537}}].

\bibitem{Henriksson:2022dml}
J.~Henriksson and B.~McPeak, \emph{{Averaging over codes and an SU(2) modular
  bootstrap}}, \href{https://doi.org/10.1007/JHEP11(2023)035}{\emph{JHEP}
  {\bfseries 11} (2023) 035}
  [\href{https://arxiv.org/abs/2208.14457}{{\ttfamily 2208.14457}}].

\bibitem{DHP}
A.~Dymarsky, J.~Henriksson and B.~McPeak, \emph{{Holographic duality from Howe
  duality: Chern–Simons gravity as an ensemble of code CFTs}},
  \href{https://arxiv.org/abs/25??.?????}{{\ttfamily 25??.?????}}

\bibitem{Gaiotto:2020iye}
D.~Gaiotto and J.~Kulp, \emph{{Orbifold groupoids}},
  \href{https://doi.org/10.1007/JHEP02(2021)132}{\emph{JHEP} {\bfseries 02}
  (2021) 132} [\href{https://arxiv.org/abs/2008.05960}{{\ttfamily
  2008.05960}}].

\bibitem{Shao:2023gho}
S.-H.~Shao, \emph{{What's Done Cannot Be Undone: TASI Lectures on
  Non-Invertible Symmetries}},
  \href{https://arxiv.org/abs/2308.00747}{{\ttfamily 2308.00747}}.

\bibitem{Kapustin:2010hk}
A.~Kapustin and N.~Saulina, \emph{{Topological boundary conditions in abelian
  Chern-Simons theory}},
  \href{https://doi.org/10.1016/j.nuclphysb.2010.12.017}{\emph{Nucl. Phys. B}
  {\bfseries 845} (2011) 393}
  [\href{https://arxiv.org/abs/1008.0654}{{\ttfamily 1008.0654}}].

\bibitem{Kapustin:2010if}
A.~Kapustin and N.~Saulina, \emph{{Surface operators in 3d Topological Field
  Theory and 2d Rational Conformal Field Theory}},
  \href{https://arxiv.org/abs/1012.0911}{{\ttfamily 1012.0911}}.

\bibitem{Roumpedakis:2022aik}
K.~Roumpedakis, S.~Seifnashri and S.-H.~Shao, \emph{{Higher Gauging and
  Non-invertible Condensation Defects}},
  \href{https://doi.org/10.1007/s00220-023-04706-9}{\emph{Commun. Math. Phys.}
  {\bfseries 401} (2023) 3043}
  [\href{https://arxiv.org/abs/2204.02407}{{\ttfamily 2204.02407}}].

\bibitem{Raeymaekers:2023ras}
J.~Raeymaekers and P.~Rossi, \emph{{Wormholes and surface defects in rational
  ensemble holography}},
  \href{https://doi.org/10.1007/JHEP01(2024)104}{\emph{JHEP} {\bfseries 01}
  (2024) 104} [\href{https://arxiv.org/abs/2312.02276}{{\ttfamily
  2312.02276}}].

\bibitem{Salton:2016qpp}
G.~Salton, B.~Swingle and M.~Walter, \emph{{Entanglement from Topology in
  Chern-Simons Theory}},
  \href{https://doi.org/10.1103/PhysRevD.95.105007}{\emph{Phys. Rev. D}
  {\bfseries 95} (2017) 105007}
  [\href{https://arxiv.org/abs/1611.01516}{{\ttfamily 1611.01516}}].

\bibitem{BDS}
A.~Barbar, A.~Dymarsky and A.D.~Shapere, \emph{{in progress}},
  \href{https://arxiv.org/abs/25??.?????}{{\ttfamily 25??.?????}}

\bibitem{hartle1976path}
J.B.~Hartle and S.W.~Hawking, \emph{Path-integral derivation of black-hole
  radiance}, {\emph{Physical Review D} {\bfseries 13} (1976) 2188}.

\bibitem{Banks:2010zn}
T.~Banks and N.~Seiberg, \emph{{Symmetries and Strings in Field Theory and
  Gravity}}, \href{https://doi.org/10.1103/PhysRevD.83.084019}{\emph{Phys. Rev.
  D} {\bfseries 83} (2011) 084019}
  [\href{https://arxiv.org/abs/1011.5120}{{\ttfamily 1011.5120}}].

\bibitem{Harlow:2018tng}
D.~Harlow and H.~Ooguri, \emph{{Symmetries in quantum field theory and quantum
  gravity}}, \href{https://doi.org/10.1007/s00220-021-04040-y}{\emph{Commun.
  Math. Phys.} {\bfseries 383} (2021) 1669}
  [\href{https://arxiv.org/abs/1810.05338}{{\ttfamily 1810.05338}}].

\bibitem{heegaard}
P.~Heegaard, \emph{Forstudier til en topologisk teori for de algebraiske
  fladers sammenhaeng}, Ph.D. thesis, Copenhagen, 1898.

\bibitem{Raeymaekers:2021ypf}
J.~Raeymaekers, \emph{{A note on ensemble holography for rational tori}},
  \href{https://doi.org/10.1007/JHEP12(2021)177}{\emph{JHEP} {\bfseries 12}
  (2021) 177} [\href{https://arxiv.org/abs/2110.08833}{{\ttfamily
  2110.08833}}].

\bibitem{Meruliya:2021lul}
V.~Meruliya and S.~Mukhi, \emph{{AdS$_{3}$ gravity and RCFT ensembles with
  multiple invariants}},
  \href{https://doi.org/10.1007/JHEP08(2021)098}{\emph{JHEP} {\bfseries 08}
  (2021) 098} [\href{https://arxiv.org/abs/2104.10178}{{\ttfamily
  2104.10178}}].

\bibitem{Gukov:2004id}
S.~Gukov, E.~Martinec, G.W.~Moore and A.~Strominger, \emph{{Chern-Simons gauge
  theory and the AdS$_3$ / CFT$_2$ correspondence}},  in \emph{{From Fields to
  Strings: Circumnavigating Theoretical Physics: A Conference in Tribute to Ian
  Kogan}}, pp.~1606--1647, 3, 2004,
  \href{https://doi.org/10.1142/9789812775344_0036}{DOI}
  [\href{https://arxiv.org/abs/hep-th/0403225}{{\ttfamily hep-th/0403225}}].

\bibitem{Freed:2022qnc}
D.S.~Freed, G.W.~Moore and C.~Teleman, \emph{{Topological symmetry in quantum
  field theory}},  \href{https://arxiv.org/abs/2209.07471}{{\ttfamily
  2209.07471}}.

\bibitem{Lin:2023uvm}
Y.-H.~Lin and S.-H.~Shao, \emph{{Bootstrapping noninvertible symmetries}},
  \href{https://doi.org/10.1103/PhysRevD.107.125025}{\emph{Phys. Rev. D}
  {\bfseries 107} (2023) 125025}
  [\href{https://arxiv.org/abs/2302.13900}{{\ttfamily 2302.13900}}].

\bibitem{Witten:1995gf}
E.~Witten, \emph{{On S duality in Abelian gauge theory}},
  \href{https://doi.org/10.1007/BF01671570}{\emph{Selecta Math.} {\bfseries 1}
  (1995) 383} [\href{https://arxiv.org/abs/hep-th/9505186}{{\ttfamily
  hep-th/9505186}}].

\bibitem{Verlinde:1995mz}
E.P.~Verlinde, \emph{{Global aspects of electric - magnetic duality}},
  \href{https://doi.org/10.1016/0550-3213(95)00431-Q}{\emph{Nucl. Phys. B}
  {\bfseries 455} (1995) 211}
  [\href{https://arxiv.org/abs/hep-th/9506011}{{\ttfamily hep-th/9506011}}].

\bibitem{Choi:2021kmx}
Y.~Choi, C.~Cordova, P.-S.~Hsin, H.T.~Lam and S.-H.~Shao, \emph{{Noninvertible
  duality defects in 3+1 dimensions}},
  \href{https://doi.org/10.1103/PhysRevD.105.125016}{\emph{Phys. Rev. D}
  {\bfseries 105} (2022) 125016}
  [\href{https://arxiv.org/abs/2111.01139}{{\ttfamily 2111.01139}}].

\bibitem{cappelli1987modular}
A.~Cappelli, C.~Otzykson and J.-B.~Zuber, \emph{Modular invariant partition
  functions in two dimensions}, {\emph{Nuclear Physics B} {\bfseries 280}
  (1987) 445}.

\bibitem{cappelli1987ade}
A.~Cappelli, C.~Itzykson and J.-B.~Zuber, \emph{The ade classification of
  minimal and a 1 (1) conformal invariant theories}, {\emph{Communications in
  Mathematical Physics} {\bfseries 113} (1987) 1}.

\bibitem{Korinman_2019}
J.~Korinman, \emph{Decomposition of some witten-reshetikhin-turaev
  representations into irreducible factors},
  \href{https://doi.org/10.3842/sigma.2019.011}{\emph{Symmetry, Integrability
  and Geometry: Methods and Applications} (2019) }.

\bibitem{balasubramanian2017multi}
V.~Balasubramanian, J.R.~Fliss, R.G.~Leigh and O.~Parrikar,
  \emph{Multi-boundary entanglement in chern-simons theory and link
  invariants}, {\emph{Journal of High Energy Physics} {\bfseries 2017} (2017)
  1}.

\bibitem{balasubramanian2018entanglement}
V.~Balasubramanian, M.~DeCross, J.~Fliss, A.~Kar, R.G.~Leigh and O.~Parrikar,
  \emph{Entanglement entropy and the colored jones polynomial}, {\emph{Journal
  of High Energy Physics} {\bfseries 2018} (2018) 1}.

\bibitem{melnikov2019topological}
D.~Melnikov, A.~Mironov, S.~Mironov, A.~Morozov and A.~Morozov, \emph{From
  topological to quantum entanglement}, {\emph{Journal of High Energy Physics}
  {\bfseries 2019} (2019) 1}.

\bibitem{PhysRevD.107.126005}
D.~Melnikov, \emph{Entanglement classification from a topological perspective},
  \href{https://doi.org/10.1103/PhysRevD.107.126005}{\emph{Phys. Rev. D}
  {\bfseries 107} (2023) 126005}.

\bibitem{Melnikov:2023wwc}
D.~Melnikov, \emph{{Connectomes as Holographic States}},
  \href{https://arxiv.org/abs/2312.16683}{{\ttfamily 2312.16683}}.

\bibitem{yahagi2022narain}
S.~Yahagi, \emph{Narain cfts and error-correcting codes on finite fields},
  {\emph{Journal of High Energy Physics} {\bfseries 2022} (2022) 1}.

\bibitem{Kawabata:2023iss}
K.~Kawabata, T.~Nishioka and T.~Okuda, \emph{{Narain CFTs from quantum codes
  and their ${\mathbb{Z}}_{2}$ gauging}},
  \href{https://doi.org/10.1007/JHEP05(2024)133}{\emph{JHEP} {\bfseries 05}
  (2024) 133} [\href{https://arxiv.org/abs/2308.01579}{{\ttfamily
  2308.01579}}].

\bibitem{benjamin2020pure}
N.~Benjamin, S.~Collier and A.~Maloney, \emph{Pure gravity and conical
  defects}, {\emph{Journal of High Energy Physics} {\bfseries 2020} (2020) 1}.

\bibitem{Maxfield:2020ale}
H.~Maxfield and G.J.~Turiaci, \emph{{The path integral of 3D gravity near
  extremality; or, JT gravity with defects as a matrix integral}},
  \href{https://doi.org/10.1007/JHEP01(2021)118}{\emph{JHEP} {\bfseries 01}
  (2021) 118} [\href{https://arxiv.org/abs/2006.11317}{{\ttfamily
  2006.11317}}].

\bibitem{PhysRevLett.132.041602}
G.~Di~Ubaldo and E.~Perlmutter, \emph{${\mathrm{ads}}_{3}$ pure gravity and
  stringy unitarity},
  \href{https://doi.org/10.1103/PhysRevLett.132.041602}{\emph{Phys. Rev. Lett.}
  {\bfseries 132} (2024) 041602}.

\bibitem{Cotler:2020ugk}
J.~Cotler and K.~Jensen, \emph{{AdS$_{3}$ gravity and random CFT}},
  \href{https://doi.org/10.1007/JHEP04(2021)033}{\emph{JHEP} {\bfseries 04}
  (2021) 033} [\href{https://arxiv.org/abs/2006.08648}{{\ttfamily
  2006.08648}}].

\bibitem{Jafferis:2024jkb}
D.L.~Jafferis, L.~Rozenberg and G.~Wong, \emph{{3d Gravity as a random
  ensemble}},  \href{https://arxiv.org/abs/2407.02649}{{\ttfamily 2407.02649}}.

\bibitem{Henriksson:2021qkt}
J.~Henriksson, A.~Kakkar and B.~McPeak, \emph{{Classical codes and chiral CFTs
  at higher genus}}, \href{https://doi.org/10.1007/JHEP05(2022)159}{\emph{JHEP}
  {\bfseries 05} (2022) 159}
  [\href{https://arxiv.org/abs/2112.05168}{{\ttfamily 2112.05168}}].

\bibitem{Henriksson:2022dnu}
J.~Henriksson, A.~Kakkar and B.~McPeak, \emph{{Narain CFTs and quantum codes at
  higher genus}}, \href{https://doi.org/10.1007/JHEP04(2023)011}{\emph{JHEP}
  {\bfseries 04} (2023) 011}
  [\href{https://arxiv.org/abs/2205.00025}{{\ttfamily 2205.00025}}].

\bibitem{gaborit1996mass}
P.~Gaborit, \emph{Mass formulas for self-dual codes over z/sub 4/and f/sub q/+
  uf/sub q/rings}, {\emph{IEEE Transactions on Information Theory} {\bfseries
  42} (1996) 1222}.

\end{thebibliography}\endgroup

\end{document}